\documentclass[sigconf]{acmart}
\renewcommand\footnotetextcopyrightpermission[1]{}
\AtBeginDocument{%
  }
\usepackage{xcolor}
\newif\ifshowrevisioncolors
\showrevisioncolorstrue

\usepackage{enumitem}
\usepackage{balance}
\usepackage[ruled,linesnumbered,vlined]{algorithm2e}
\usepackage{amsmath,amssymb}
\usepackage{microtype}           
\newcommand{\Ccur}{C_{\mathrm{cur}}}
\newcommand{\Ctgt}{C_{\mathrm{tgt}}}
\newcommand{\Ctmp}{C_{\mathrm{int}}}
\newcommand{\Madd}{\mathcal{M}_{\mathrm{add}}}
\newcommand{\Mdel}{\mathcal{M}_{\mathrm{del}}}
\newcommand{\Mmig}{\mathcal{M}_{\mathrm{mig}}}
\newcommand{\Bmax}{B_{\max}}
\newcommand{\Bcur}{B_{\mathrm{cur}}}
\newcommand{\Bnew}{B_{\mathrm{new}}}
\usepackage{graphicx} % 确保图片能够被正确插入 
\usepackage{float} % 提供 H 选项
\usepackage{titlesec}
\usepackage{booktabs}
\usepackage{amsmath}

\begin{document}

\title[]{PipeLive: Efficient Live In-place Pipeline Parallelism Reconfiguration for Dynamic LLM Serving}
% Reinforcement Learning-based
% Rescheduling of Microservice
% Architecture Applications in the
% Cloud-Edge Continuum

% If you give a name to your approach it would be more catchy!! 

% For example: REALM (REinforcement-based Adaptive rescheduling for cLoud-edge Microservices)

% RACE (Reinforcement for Application Continuum Execution)

% MERLIN (Microservice Edge Rescheduling with Learning-based Intelligence for Networks)

% or ReSchedIt
% FlexiSched
\author{Xu BAI}
\email{bxb1@student.unimelb.edu.au}
\affiliation{%
  \institution{DisNet Lab}
  \institution{University of Melbourne}
  \city{Melbourne}
  \state{VIC}
  \country{Australia}
}
\author{Muhammed Tawfiqul Islam}
\email{tawfiqul.islam@unimelb.edu.au}
\affiliation{%
  \institution{DisNet Lab}
  \institution{University of Melbourne}
  \city{Melbourne}
  \state{VIC}
  \country{Australia}
}
\author{Chen Wang}
\email{Chen.Wang1@ibm.com}
\affiliation{%
  \institution{IBM Research}
  \city{Yorktown Heights}
  \state{NY}
  \country{USA}
}
\author{Adel N. Toosi }
\email{adel.toosi@unimelb.edu.au}
\affiliation{%
  \institution{DisNet Lab}
  \institution{University of Melbourne}
  \city{Melbourne}
  \state{VIC}
  \country{Australia}
}

\begin{abstract}
%Efficiently serving large language models (LLMs) under constrained GPU resources has become a central systems challenge. 
Pipeline parallelism (PP) is widely used to partition layers of large language models (LLMs) across GPUs, enabling scalable inference for large models. However, existing systems rely on static PP configurations that fail to adapt to dynamic settings, such as serverless platforms and heterogeneous GPU environments. Reconfiguring PP by stopping and redeploying service incurs prohibitive downtime, so reconfiguration must instead proceed live and in place, without interrupting inference. However, live in-place PP reconfiguration is fundamentally challenging. GPUs are already saturated with model weights and KV cache, leaving little room for new layer placements and necessitating KV cache resizing, at odds with systems like vLLM that preallocate for throughput. Moreover, maintaining KV consistency during execution is difficult: stop-and-copy introduces large pauses, while background synchronization risks inconsistency as states evolve. We present \textsc{PipeLive}, which enables live in-place PP reconfiguration with minimal disruption. PipeLive introduces a redesigned KV cache layout together with a co-designed extension to PageAttention, forming a unified mechanism for live KV resizing. It further adopts an incremental KV patching mechanism, inspired by live virtual machine migration, to synchronize KV states between source and target configurations and identify a safe switch point. PipeLive achieves a 2.5× reduction in time-to-first-token (TTFT) without KV cache overflow. Furthermore, compared to a variant without KV patching, it reduces reconfiguration overhead to under 10ms. On a real-world trace, \textsc{PipeLive} reduces mean TPOT, TTFT, and end-to-end latency by 14.65\%, 10.04\%, and 8.65\%, respectively, relative to the best static PP configuration; it further reduces P99 TPOT and TTFT by 33.64\% and 22.91\%.

\end{abstract}



\keywords{Large Language Model, Model Serving, Pipeline Parallelism}

\maketitle

\section{Introduction}

Large language models (LLMs)~\cite{openai2023gpt4, touvron2023llama, brown2020gpt3} have advanced rapidly in recent years. Due to their large parameter sizes and high computational demands, serving a single model often requires coordinating multiple GPUs. \textit{Pipeline parallelism} (PP)~\cite{narayanan2019pipedream, shoeybi2020megatronlm} is a widely adopted strategy that partitions model layers across GPUs to scale inference, and the choice of a PP configuration---the pipeline depth and the layer-to-GPU assignment---that defines the partitioning of model layers across GPUs significantly affects system performance~\cite{zheng2023alpa}. Existing LLM inference systems~\cite{kwon2023vllm, zheng2024sglang, he2025uellm, yu2022orca} treat this configuration as a static decision fixed at deployment time: they are optimized for throughput with preallocated model weights and KV cache, so changing the layer-to-GPU mapping requires a full service restart that incur minutes-long downtime, discards in-flight computation, and operate at a timescale fundamentally mismatched with second-level workload dynamics~\cite{fu2024serverlessllm}.

In practice, neither the resources nor the workload that a serving system faces stay constant. Request rates fluctuate over minutes, so an elastic deployment must scale the pipeline itself---deepening it when additional GPUs are allocated to absorb load spikes, and shrinking it when the cluster reclaims capacity or serverless platforms withdraw GPUs at runtime. This need for elasticity arises even in homogeneous settings: with identical GPUs, redistributing model layers across a varying number of devices is what enables autoscaling in response to demand~\cite{lin2025flexpipe}. Heterogeneous environments add a second axis of adaptation. Even at a fixed depth, the request mix shifts over time between prefill-heavy and decode-heavy regimes~\cite{wang2024burstgpt}, pulling the ideal layer partition in opposite directions: long-prompt requests are compute-bound in attention and favor placing more layers on compute-strong GPUs, whereas decode-dominated requests are memory-bandwidth-bound from frequent KV cache accesses and favor high-bandwidth GPUs~\cite{suSeesawHighthroughputLLM2025}. Today, both forms of adaptation---changing the pipeline depth and re-balancing the partition---require a restart whose disruption outweighs the misconfiguration it fixes. Live, in-place PP reconfiguration removes this barrier: it lets the system switch between workload-optimal configurations at runtime, growing, shrinking, or re-partitioning the pipeline with minimal disruption and without borrowing spare GPUs that constrained clusters do not have.

Achieving such live, in-place PP reconfiguration is practically challenging due to several key factors.
% First, in-place PP reconfiguration must redistribute KV cache and model weights on GPUs that are actively serving inference. The system must keep the source configuration running while preparing the target configuration, leading to temporary coexistence of weights and KV states from both configurations. To support this coexistence, a robust reconfiguration protocol is required to dynamically manage KV memory. This includes shrinking KV cache without disrupting ongoing inference to make room for incoming layers, asynchronously loading new weights, coordinating KV state transfer across GPUs, and determining a safe switchover point to complete reconfiguration.
First, the system must keep the initial configuration running while preparing the target configuration, leading to the temporary coexistence of model weights and KV states from both configurations. Existing state-of-the-art systems~\cite{kwon2023vllm,zheng2024sglang} typically preallocate GPU memory for KV cache and weights, leaving no room for dynamically reclaiming KV cache memory to accommodate the weights and KV states of newly loaded layers, thereby rendering in-place PP reconfiguration infeasible.

Second, in-place PP reconfiguration requires resizing the live KV cache, which is fundamentally constrained by the contiguous memory layout used in existing systems. State-of-the-art LLM serving systems~\cite{kwon2023vllm,zheng2024sglang} adopt PageAttention~\cite{kwon2023vllm}, which organizes KV cache into fixed-size blocks but preallocates each layer’s KV cache as a contiguous region in GPU memory for optimized inference throughput. This fundamentally precludes dynamic resizing, as GPU memory cannot be resized in place. A natural approach to enable resizing is to adopt block-level KV allocation, allowing KV cache to grow or shrink by allocating or releasing blocks at runtime. However, GPU memory is allocated at a much coarser granularity than typical KV blocks, making naive block-level allocation prohibitively inefficient due to severe internal fragmentation. Together, these issues make efficient KV cache resizing during reconfiguration fundamentally challenging.

Third, PP reconfiguration must preserve KV state consistency without completely blocking the incoming request or ongoing model inference. %across initial and target configurations during concurrent execution.
Because KV cache migration volumes are large, stop-and-copy transfers incur prohibitive pauses, while in-flight transfers will sacrifice consistency in the KV Cache. The challenge is to ensure KV consistency without introducing synchronization overhead.

% \textsc{PipeLive} adopts a hierarchical design that separates global coordination from per-GPU execution. 

Taken together, these challenges make live, in-place PP reconfiguration non-trivial to achieve in practice. To address these challenges, we present \textsc{PipeLive}, an open-source LLM serving system that enables efficient and low-disruption live, in-place PP reconfiguration, allowing systems to adapt to dynamic workloads at runtime. To handle the complex KV cache resizing and migration on GPUs during in-place reconfiguration, we design a robust \textbf{coordination protocol} that orchestrates these operations, ensuring the safe coexistence of source and target configurations. To support efficient KV resizing, we extend PageAttention with \textbf{non-contiguous KV block access} to enable block-level KV allocation, and introduce \textbf{layer stacking} to pack multiple layers’ KV states into a single allocation unit, thereby aligning logical KV blocks with GPU allocation granularity and mitigating internal fragmentation. To minimize disruption, we further introduce a \textbf{KV patching} mechanism, inspired by VM live migration~\cite{clark2005live}, which incrementally migrates KV states between source and target layers during inference, reducing divergence to a small residual that can be reconciled with a short final pause.
% Our \textbf{main contributions} are as follows:
% \begin{itemize}
%     \item We present \textsc{PipeLive}, a unified system that integrates a set of carefully designed system enhancements to enable efficient live, in-place PP reconfiguration, built on top of a widely used open-source LLM serving framework, vLLM.

%     \item We design a robust coordination protocol that orchestrates KV cache resizing and migration, as well as model weight loading, across multiple GPUs, enabling in-place PP reconfiguration to proceed concurrently and efficiently.
    
%     \item We design a dynamic KV cache management and migration mechanism for in-place PP reconfiguration. We extend PageAttention to support block-level KV allocation and introduce a layer-stacking scheme to align logical KV blocks with GPU allocation granularity, enabling efficient resizing while mitigating internal fragmentation. We further introduce a continuous KV patching mechanism to migrate layers with minimal disruption while preserving KV consistency.

 We summarize our \textbf{main contributions} as follows:
\begin{itemize}
    \item We present \textsc{PipeLive}, a unified system for live, in-place PP reconfiguration, built on top of the widely used open-source LLM serving framework vLLM~\cite{kwon2023vllm}. 

    \item We design a robust coordination protocol to enable safe and efficient in-place PP reconfiguration.

    \item We develop a dynamic KV cache management mechanism for reconfiguration by extending PagedAttention with non-contiguous KV block access and introducing a layer-stacking technique for efficient KV resizing.

    \item We introduce an incremental KV patching mechanism that preserves KV consistency and minimizes service disruption during PP reconfiguration.
\end{itemize}
    
    % \item We conduct comprehensive evaluations to isolate the impact of each optimization. Results show that \textsc{PipeLive} enables efficient switching between workload-dependent PP configurations in heterogeneous GPU environments, outperforming static deployments, while maintaining low reconfiguration overhead and reducing service interruption to the millisecond scale.
%\end{itemize}
% Backup
% \begin{figure}
%     \centering
%     \includegraphics[width=0.5\linewidth]{Figures/ce-arch.pdf}
%     \caption{Architecture of Cloud-Edge Continuum}
%     \label{fig:intro-ce}
% \end{figure}

\section{Background and Motivation}
\subsection{LLM Inference}

Compared to conventional AI workloads, LLM inference operates on variable-length inputs and produces variable-length outputs~\cite{yuanLLMInferenceUnveiled2024}. Prompt processing is referred to as the \emph{prefill} stage: the model consumes the prompt and computes activations for all prompt tokens in parallel, and this stage is typically compute-bound. After prefill, the model generates one token at a time and feeds each token back as input, producing an output sequence autoregressively; this process is the \emph{decode} stage. Decoding is often memory-bound due to its low arithmetic intensity and repeated access to model weights and cached attention states. These contrasting resource profiles motivate prefill--decode (PD) disaggregation, which places the two phases on separate GPU pools to avoid cross-phase interference and independently optimize resource allocation and parallelism for each phase, at the cost of transferring each request's KV state between pools~\cite{zhong2024distserve,qinMooncake2025}.

During inference, LLMs materialize key–value (KV) caches to accelerate subsequent token generation~\cite{yu2022orca}. As context lengths and request concurrency increase, KV cache memory often becomes the dominant memory footprint~\cite{qinMooncake2025}. \emph{PagedAttention}~\cite{kwon2023vllm} improves KV cache utilization by managing the cache in fixed-size blocks (pages) and allocating these blocks to requests on demand, thus reducing memory fragmentation.

LLM serving performance is typically characterized by three metrics. \emph{Throughput} measures steady-state generation capacity (e.g., output tokens or requests per second) and is central for batch-style workloads. \emph{Time-to-first-token (TTFT)} captures the latency from request arrival to the first generated token. \emph{Time-per-output-token (TPOT)} measures the average latency per generated token after the first token. Different workload patterns emphasize different metrics, creating inherent trade-offs in the design of serving systems.

\subsection{Deployment of LLM Models}
Pipeline parallelism is a widely adopted approach in which model layers are partitioned across GPUs~\cite{shoeybi2020megatronlm}; during inference, input activations flow sequentially through the pipeline, with each GPU computing its assigned layers before forwarding intermediate results to the next stage.

Most existing LLM serving systems~\cite{kwon2023vllm, zheng2024sglang, he2025uellm, yu2022orca} adopt a static PP configuration, in which a fixed PP configuration is determined via offline profiling before system launch and remains unchanged throughout the serving lifecycle. Although simple, this approach creates a fundamental mismatch between allocated GPU resources and evolving LLM workloads.

\subsection{Motivation}
\label{sec:motivation}
\begin{figure}
    \centering
    \includegraphics[width=0.85\columnwidth]{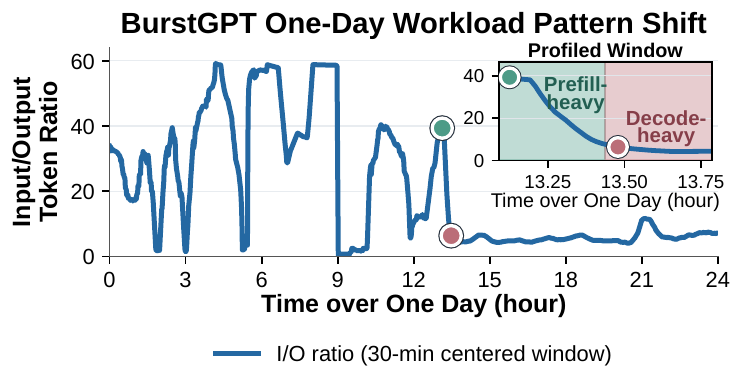}
    \caption{BurstGPT one-day input/output token-ratio trace using a 30-minute centered window. The marked intervals identify the profiled prefill-heavy and decode-heavy workloads.}
    \label{fig:motivation-workload}
\end{figure}

\begin{figure}
    \centering
    \includegraphics[width=0.85\columnwidth]{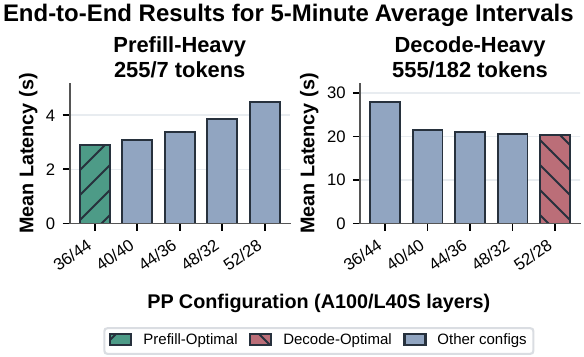}
    \vspace{-0.35em}
    \caption{Mean end-to-end latency under the prefill-heavy and decode-heavy workloads across PP configurations on the A100--L40S testbed.}
    \label{fig:motivation-1}
\end{figure}

{
Existing serving systems treat the PP configuration (the pipeline depth and the layer-to-GPU assignment) as a static decision fixed at deployment time. This assumption breaks down in the increasingly dynamic environments where LLM inference is deployed today. On cloud platforms using preemptible GPUs, instance availability changes at runtime~\cite{miao2024spotserve}; in resource-constrained deployments, GPU servers may be loaned from inference to training during low-traffic periods and reclaimed as inference demand rises~\cite{li2023lyra}. In both cases, the serving deployment must adapt as its available capacity changes. Moreover, even when resources remain fixed, the optimal configuration itself drifts: the layer partition that balances the pipeline depends on workload characteristics, which shift continuously in production. A serving system that cannot adapt its PP configuration must therefore either overprovision for the worst case or operate persistently far from the optimum.

Stopping and redeploying the service is not a practical answer. Reloading tens to hundreds of gigabytes of model weights, discarding in-flight requests, and rebuilding KV-cache state can introduce minutes of downtime and violate latency SLOs. Reconfiguration must instead be live, preserving ongoing inference, and in-place, without relying on additional GPU resources. The conventional cloud approach of deploying the new configuration on spare GPUs before redirecting traffic assumes idle capacity that resource-constrained clusters~\cite{lin2025flexpipe,fu2024serverlessllm}, by definition, lack. The new layer placement must instead be realized directly on GPUs already occupied by model weights and preallocated KV cache. Live, in-place PP reconfiguration is thus not merely a performance optimization; it is a fundamental requirement for elastic and adaptive LLM serving.

One setting where such reconfiguration is particularly valuable---and which serves as the evaluation setting for this work---is heterogeneous pipeline parallelism, in which a single LLM is partitioned across GPUs with different compute and memory capabilities. Unlike prefill--decode (PD) disaggregation, which replicates the full model across separate prefill and decode pools and incurs KV-state transfer overhead~\cite{zhong2024distserve}, heterogeneous pipeline parallelism maintains a single distributed copy of the model, making it attractive for resource-constrained and on-premises deployments~\cite{moHetis2025}. The optimal partition, however, is not fixed: it depends on the computational capabilities of the participating GPUs as well as workload characteristics such as request arrival rates and input-to-output token ratios. Heterogeneous pipeline parallelism can therefore realize its full potential only if the pipeline can be reconfigured dynamically without interrupting ongoing inference.

Figures~\ref{fig:motivation-workload} and~\ref{fig:motivation-1} illustrate this phenomenon using BurstGPT~\cite{wang2024burstgpt}. The workload trace highlights a consecutive 30-minute intervals that transition from a prefill-heavy regime with average input/output lengths of $255/7$ to a decode-heavy regime with average lengths of $555/182$. We replay both regimes on a two-stage A100--L40S pipeline serving the 80-layer Llama-70B model. As the workload shifts between regimes, the relative performance of the two stages changes, and a different layer partition is required to keep the pipeline balanced. The end-to-end results confirm this: the optimal partition moves from $36/44$ in the prefill-heavy regime to $52/28$ in the decode-heavy regime, and a partition optimized for one regime incurs substantially higher latency in the other. No single static partition can serve both regimes efficiently.

These observations motivate PipeLive, a general mechanism for live, low-overhead, in-place PP reconfiguration. PipeLive relocates layers across running GPUs without stopping or redeploying the model, keeping the reconfiguration cost low enough that adapting the pipeline pays off in practice.}

\section{Overview of \textsc{PipeLive}}

\begin{figure}[t]
    \centering
    \includegraphics[width=0.9\columnwidth]{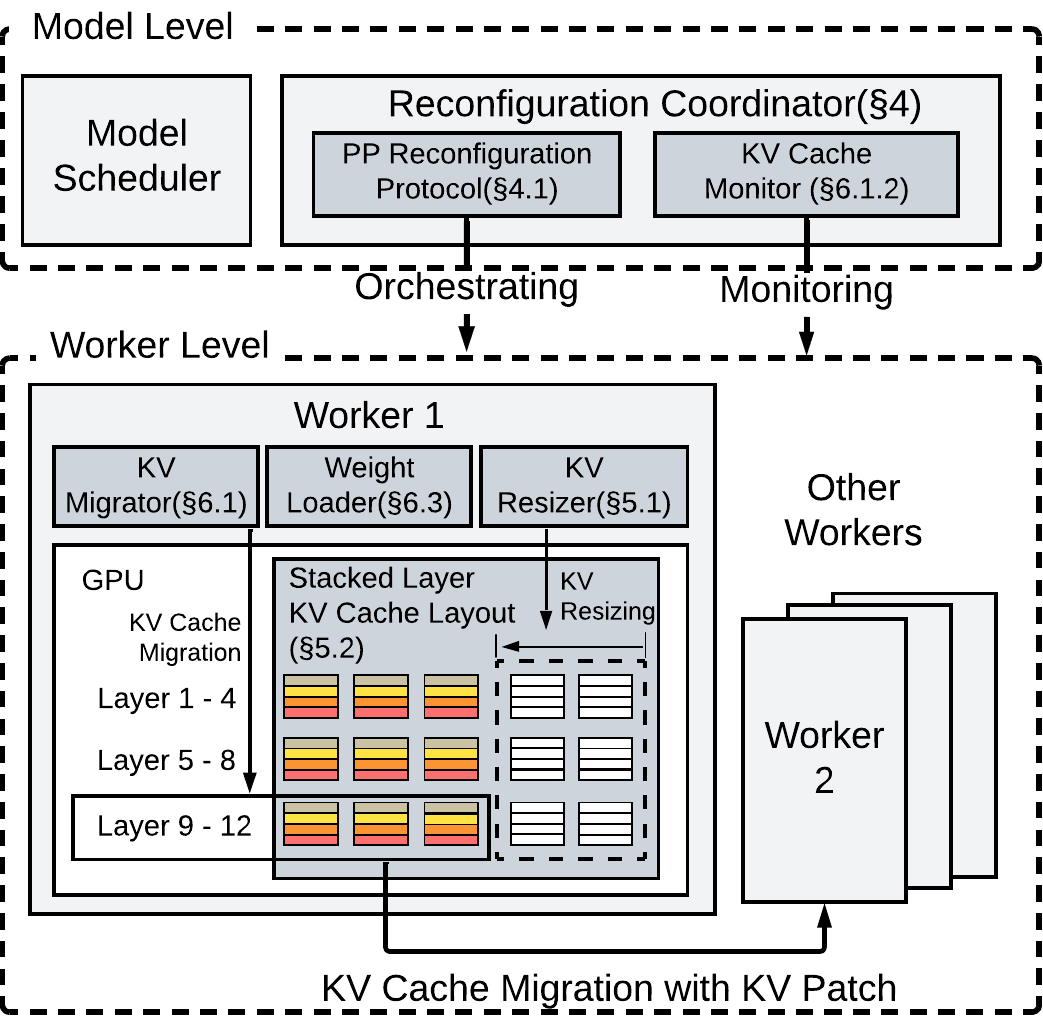}
  \caption{Architecture of \textsc{PipeLive}.}
    \label{fig:flexiserve-arch}
\end{figure}

\textsc{PipeLive} adopts a top-down design to enable efficient in-place PP reconfiguration. Figure~\ref{fig:flexiserve-arch} illustrates the overall architecture. 

At the \emph{model level}, alongside the request scheduler that handles incoming requests and dispatches them to GPUs, \textsc{PipeLive} introduces a centralized \emph{Reconfiguration Coordinator}. Given the current and target PP configurations, the coordinator executes a reconfiguration protocol that (1) evaluates feasibility under GPU memory constraints and KV cache state, and (2) synthesizes an explicit execution plan that decomposes the target configuration into coordinated actions, including \textit{KV cache resizing} and \textit{migration}, and \textit{asynchronous weight loading}. During execution, the coordinator leverages a KV Cache Monitor to continuously track KV migration progress across GPUs and determine a safe switchover point for committing the new configuration, minimizing disruption to ongoing inference.

At the \emph{worker level}, each device runs a local \emph{Reconfiguration Worker} with three components: (1) A \emph{KV cache resizer} that adjusts KV capacity to match layer placement, (2) A \emph{weight loader} that keeps weights in CPU memory and stages newly needed layers to the target GPU on demand, and (3) A \emph{KV cache migrator} that transfers KV states between the source and destination GPUs while preserving correctness under ongoing inference. To minimize disruption, the migrator continuously patches newly generated KV states, reducing migration overhead to a short final cutover.

Furthermore, \textsc{PipeLive} extends PagedAttention with a block-level KV cache preallocation strategy. This design enables efficient KV cache resizing and compaction at block granularity while retaining performance comparable to the original PagedAttention kernel. To align the GPU’s minimum physical allocation granularity with logical KV cache blocks, we introduce a \emph{layer stacking} technique that packs KV states from multiple layers into a single allocation unit. This co-design of memory layout and allocation granularity improves utilization and mitigates fragmentation during dynamic resizing.

% PipeLive's design enables efficient PP reconfiguration with minimal disruption to ongoing inference. By coordinating reconfiguration across the model and worker levels, and by optimizing KV cache management for dynamic resizing and migration, \textsc{PipeLive} minimizes downtime and bounds the impact of reconfiguration on key performance metrics, enabling practical dynamic PP reconfiguration for LLM serving under heterogeneous and evolving workloads.

In the following sections, we first describe the reconfiguration worker in \S\ref{sec:reconfiguration_coordinator}, then present the KV cache resizer and layer stacking in \S\ref{sec:kv_cache_management}, and finally introduce the KV migrator, along with its KV patching mechanism and model weight loader, in \S\ref{sec:state_sync}.

\section{Reconfiguration Coordinator}
\label{sec:reconfiguration_coordinator}
The Reconfiguration Coordinator is the central component that orchestrates PP reconfiguration in \textsc{PipeLive}. It implements a PP reconfiguration protocol that defines how a target PP configuration is realized through a sequence of coordinated actions across GPUs.  In this section, we first present an end-to-end overview of the PP reconfiguration workflow. We then describe the core reconfiguration primitives defined by the PP reconfiguration protocol and executed by the Reconfiguration Coordinator, along with their semantics. Finally, we introduce a reconfiguration algorithm that composes these primitives to safely and efficiently perform live in-place PP reconfiguration at runtime.

% Rather than treating reconfiguration as a monolithic operation, the protocol decomposes PP reconfiguration into a set of well-defined \emph{primitives}, each capturing a fundamental reconfiguration action. These primitives are invoked via collective RPCs and executed by GPU-level Reconfiguration Worker in a coordinated manner, enabling reconfiguration to proceed concurrently with inference while tightly controlling resource usage and interference.

\subsection{General Reconfiguration Procedure}
\begin{figure}[]
    \centering
    \includegraphics[width=1\columnwidth]{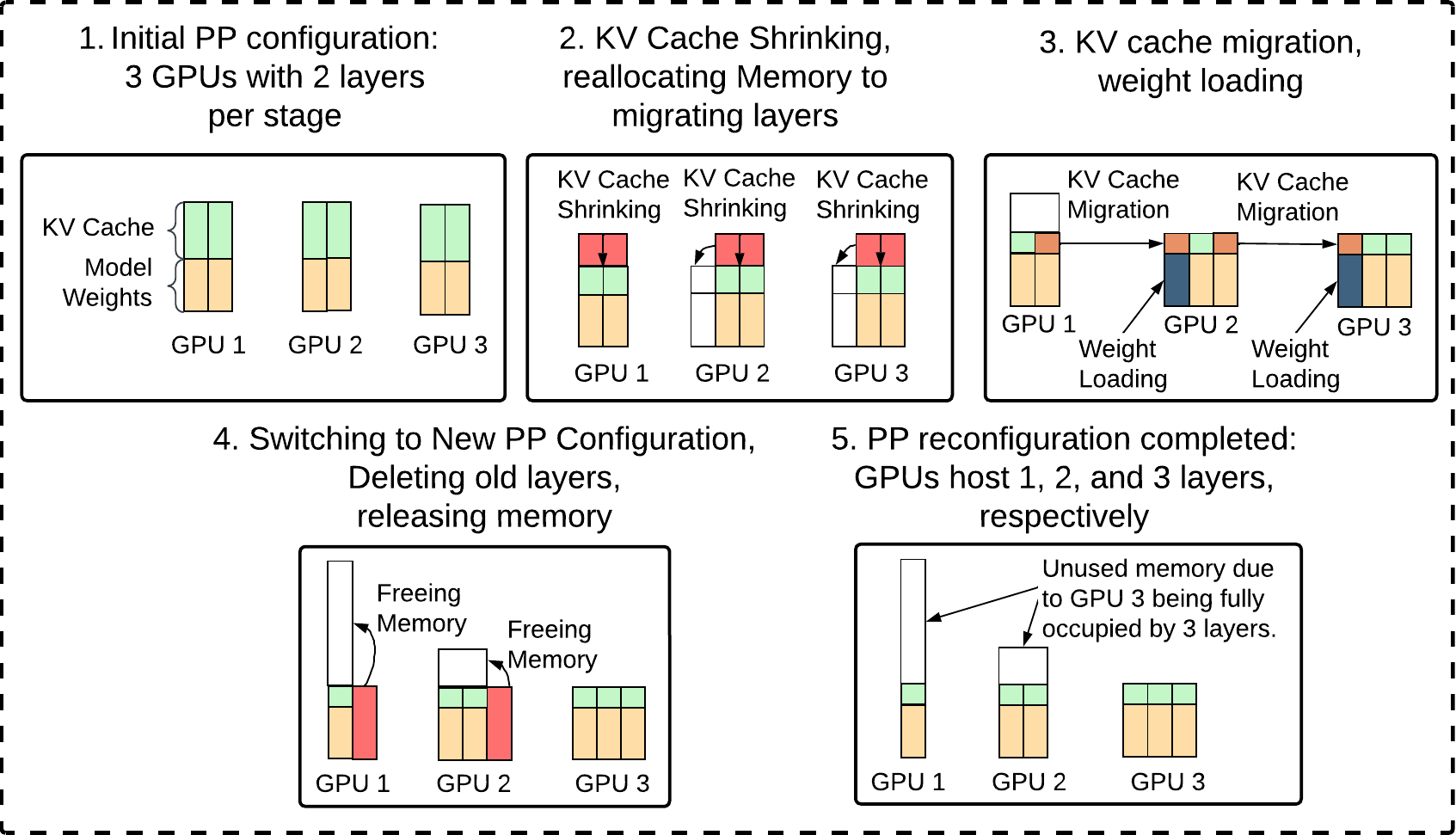}
  \caption{Live in-place PP reconfiguration workflow in \textsc{PipeLive}.}
    \label{fig:flexiserve-procedure}
\end{figure}

Figure~\ref{fig:flexiserve-procedure} illustrates a canonical PP reconfiguration workflow. We represent a PP configuration as an ordered tuple of layer ranges assigned to each GPU. For clarity, we consider a simplified setup with three GPUs (GPU~1, GPU~2, GPU~3) serving a six-layer model. 

\textbf{1. Initial PP Configuration.} The system transitions from an initial configuration $\mathcal{C}_A = \langle [1,2],[3,4],[5,6] \rangle$ to a target configuration $\mathcal{C}_B = \langle [1],[2,3],[4,5,6] \rangle$, requiring layer migration across adjacent GPUs.

\textbf{2. KV Cache Shrinking.} At the outset, GPU memory is largely occupied by model weights and the KV cache. To admit incoming layers, all GPUs first compute the required space for both the weights and KV states of migrated layers according to the protocol in \S\ref{sec:protocal}, and then shrink their KV cache accordingly. Notably, we apply KV cache shrinking to all GPUs, including those that neither load new weights nor receive KV states (e.g., GPU1), to maintain consistent KV cache capacity across all layers.

\textbf{3. KV Cache Migration.} Reconfiguration then proceeds by overlapping weight loading and KV migration. Destination GPUs preload the incoming layer weights, while source GPUs concurrently stream KV states to their new locations. In this setting, GPU~2 simultaneously receives and forwards layers; \textsc{PipeLive} executes these transfers asynchronously, enabling concurrent send/receive to accelerate reconfiguration. Meanwhile, inference continues under $\mathcal{C}_A$ during this phase. As execution progresses, newly generated KV entries are incrementally synchronized to destination GPUs via KV patching (\S~\ref{sec:kv-synchronizer}), preventing large synchronization stalls.

\textbf{4. Switching PP Configuration.} The coordinator tracks migration progress using KV indices at sources and destinations. Once the lag falls below a threshold, it triggers a brief synchronization point, performs a final KV transfer, and atomically switches to $\mathcal{C}_B$. After the switch, source GPUs release migrated layers and reclaim memory. If, after PP reconfiguration, there is reclaimable GPU memory available for the KV cache, the KV cache will resize to restore usable capacity. However, in this case, after migration, no more KV cache is reclaimable as GPU~3 is full.

During live in-place PP reconfiguration, each GPU performs a distinct sequence of actions determined by the source and target PP configurations. We design a PP reconfiguration protocol that formalizes the required primitives and a reconfiguration algorithm that composes these primitives. Note that this work focuses on efficiently executing a requested live in-place PP reconfiguration, with minimal overhead and disruption to LLM inference. We do not address when or why reconfiguration should be triggered, nor how to determine the optimal source and target PP configurations, and these aspects are therefore deferred to future work.

% must correctly and concurrently coordinate the sequence of stateful, latency-sensitive actions and commit the configuration change consistently across all devices.

\subsection{Pipeline Parallelism Reconfiguration Protocol}
\label{sec:protocal}
The PP reconfiguration protocol exposes a set of \emph{primitives} that the Reconfiguration Coordinator invokes to orchestrate coordinated reconfiguration across GPUs. Based on their invocation semantics, these primitives are categorized into two types: \emph{Collective Primitives (CP)} and \emph{Synchronization Primitives (SP)}. CPs are issued by the Reconfiguration Coordinator via collective Remote Procedure calls (RPCs) and are asynchronously executed by all Reconfiguration Workers. In contrast, SPs are embedded within model scheduler requests and are sequentially executed by GPUs in pipeline order, thereby introducing explicit synchronization points into the inference process.

To uniformly represent per-GPU actions in collective primitives, we use a
\emph{layer assignment map} $\mathcal{M}$. $\mathcal{M}$ maps a GPU index to a set of
layers, i.e.,
$\mathcal{M} : i \mapsto \{\ell^{(i)}_{1}, \ell^{(i)}_{2}, \dots\}$,
where $\mathcal{M}(i)$ specifies the layer set on which GPU $i$ should operate
for a given primitive. For primitives involving data transfer across GPUs, we
use a directional assignment map $\mathcal{M}_{\text{mig}}$, where
$\mathcal{M}_{\text{mig}}(i_s, i_d)$ specifies the set of layers to be transferred from source GPU $i_s$ to destination GPU $i_d$.
The primitives are defined as follows:

\begin{itemize}

    % \item \textbf{\textsc{Collective::CompactKV}$(B)$}:  
    % Compacts the KV cache on each GPU to $B$ blocks, reclaiming fragmented space required for subsequent KV resizing operation.
    
    \item \textbf{\textsc{Collective::ResizeKV}$(B)$}:  
    Resizes the KV cache capacity on each GPU to $B$ blocks. If $B$ exceeds the current number of blocks, the cache is expanded; otherwise, it is shrunk.
    \item \textbf{\textsc{Collective::AddLayerWeights}$(\mathcal{M}_{\text{add}})$}:  
    Instructs each GPU $i$ to asynchronously load the weights
    for the layers specified in $\mathcal{M}_{\text{add}}(i)$. 

    \item \textbf{\textsc{Collective::StartKVMigration}$(\mathcal{M}_{\text{mig}})$}:  
    Initiates asynchronous KV cache migration according to
    $\mathcal{M}_{\text{mig}}$, where $\mathcal{M}_{\text{mig}}(i_s, i_d)$ specifies
    the layers whose KV states should be streamed from source GPU $i_s$ to
    destination GPU $i_d$.

    \item \textbf{\textsc{Sync::SyncAndCommit}$(\mathcal{M}_{\text{del}},\, B_{\text{new}})$}:  
    Performs a synchronized state transition that atomically commits the new PP
    configuration. After the commit, each Reconfiguration Worker on GPU $i$ deletes the
    layer weights and KV cache entries specified in $\mathcal{M}_{\text{del}}(i)$,
    and resizes the KV cache to $B_{\text{new}}$ blocks to reclaim freed memory.

\end{itemize}

\begin{table}[t]
\scriptsize
\caption{Notations}
\label{tab:notation}
\centering
\begin{tabular}{@{}ll@{}}
\toprule
\textbf{Symbol} & \textbf{Description} \\
\midrule
$\Ccur,\Ctgt,\Ctmp$ & $i \mapsto \{\ell_1,\dots\}$: GPU $\to$ layer set (current / target / intermediate) \\
$\Madd,\Mdel$ & $i \mapsto \{\ell_1,\dots\}$: GPU $\to$ layers to add / delete \\
$\Mmig$ & $i_{src} \mapsto (i_{dst} \mapsto \{\ell_1,\dots\})$: layers to migrate \\
$\Bmax(i)$ & Max KV blocks GPU $i$ can hold \\
$\Bcur,\Bnew$ & Current / post-reconfiguration KV block budget \\
$\mathcal{G}$ & Set of all GPUs, $\mathcal{G} = \{0, 1, \dots, N{-}1\}$ \\
$i$ & GPU index, $i \in \mathcal{G}$ \\
$N$ & Total number of GPUs, $N = |\mathcal{G}|$ \\
$M_i$ & Total GPU memory on GPU $i$ \\
$u$ & KV cache utilization ratio, $u \in (0,1]$ \\
$W$ & Weight memory per layer \\
$P$ & KV page (block) size \\
$\tau$ & Convergence threshold (tokens) \\
$\Delta t$ & Polling interval \\
\bottomrule
\end{tabular}
\end{table}

\begin{algorithm}[tb]
\fontsize{8}{8.3}\selectfont
\setlength{\algomargin}{0.7em}
\SetAlgoNlRelativeSize{-2}
\SetInd{0.35em}{0.55em}
\SetSideCommentLeft
\DontPrintSemicolon
\SetAlgoLined
\SetKwInput{KwIn}{Input}
\SetKwInput{KwOut}{Output}
\SetKwFunction{CompactKV}{Collective::CompactKV}
\SetKwFunction{ResizeKV}{Collective::ResizeKV}
\SetKwFunction{AddWeights}{Collective::AddLayerWeights}
\SetKwFunction{StartMig}{Collective::StartKVMigration}
\SetKwFunction{SyncCommit}{Sync::SyncAndCommit}
\SetKwFunction{MaxBlocks}{MaxBlocks}
\SetKwFunction{GetAppliedToken}{Collective::GetLastSyncedTokenIndex}
\SetKwFunction{GetSchedToken}{getLastScheduledTokenIndex}
\SetKwComment{cmt}{$\triangleright$\ }{}

\SetKwProg{Fn}{Function}{:}{}
\Fn{\MaxBlocks{$i, L$}}{
  \Return $\lfloor (M_i \cdot u - L \cdot W)/(L \cdot P) \rfloor$
}

\KwIn{$\Ccur = \{(s_i, e_i)\}_{i=0}^{N-1}$: current PP config, GPU $i$ owns layers $[s_i, e_i]$;
  $\Ctgt = \{(s'_i, e'_i)\}_{i=0}^{N-1}$: target PP config;
  $\tau$: convergence threshold (tokens)}
\KwOut{$\mathrm{true}$ if reconfiguration succeeds; $\mathrm{false}$ if infeasible}

\BlankLine
%% Phase 1
\cmt*[]{{\bfseries Phase~1: Feasibility assessment \& intermediate configuration}}

$\Ctmp \gets \Ccur$;\quad
$\Madd \gets \emptyset$\;

\ForEach(\cmt*[f]{build intermediate config $\Ctmp$}){GPU $i \in \mathcal{G}$}{
  $\Ctmp[i] \gets \Ccur[i] \cup \Ctgt[i]$
  \cmt*[f]{new layers GPU $i$ must add}\;
  $\Bmax(i) \gets$ \MaxBlocks{$i,\, |\Ctmp[i]|$}\;
}
$B_{\mathrm{shrink}} \gets \min_{i \in \mathcal{G}}\, \Bmax(i)$\;

$B_{\mathrm{used}} \gets$ current number of KV blocks in use\;
\If{$B_{\mathrm{used}} > B_{\mathrm{shrink}}$}{
  \Return $\mathrm{false}$
  \cmt*[f]{insufficient memory for reconfiguration, abort}\;
}
\BlankLine
%% Phase 2
\cmt*[]{{\bfseries Phase~2: KV Resizing}};
\If{$B_{\mathrm{shrink}} < \Bcur$}{
  % \CompactKV{$B_{\mathrm{shrink}}$}\;
  \ResizeKV{$B_{\mathrm{shrink}}$}
}
\BlankLine
%% Phase 3
\cmt*[]{{\bfseries Phase~3: Asynchronous weight loading and KV Cache Migration}(non-blocking, concurrent with inference)}

\AddWeights{$\Madd$}\;

$\Mmig \gets \emptyset$\;
\ForEach{$(i_{dst}, A_{i_{dst}}) \in \Madd$}{
  \ForEach{layer $\ell \in A_{i_{dst}}$}{
    $i_{src} \gets$ GPU owning layer $\ell$ in $\Ccur$\;
    $\Mmig[i_{src}][i_{dst}] \gets \Mmig[i_{src}][i_{dst}] \cup \{\ell\}$
  }
}
\StartMig{$\Mmig$}

\BlankLine
%% Phase 4
\cmt*[]{{\bfseries Phase~4: Convergence monitoring}};
\Repeat{}{
  $T_{\mathrm{sched}} \gets$ \GetSchedToken{}\;
  $T_{\mathrm{applied}}[\,] \gets$ \GetAppliedToken{}\;
  $\mathit{allDone} \gets \mathrm{true}$\;
  \ForEach{GPU $i \in \mathrm{dom}(\Madd)$}{
    \If{$T_{\mathrm{sched}} - T_{\mathrm{applied}}[i] \ge \tau$}{
      $\mathit{allDone} \gets \mathrm{false}$\;
      \textbf{break}\;
    }
  }
  \lIf{$\mathit{allDone}$}{\textbf{break}}
  \textbf{sleep}($\Delta t$)\;
}

\BlankLine
%% Phase 5
\cmt*[]{{\bfseries Phase~5: Commit PP Configuration Change}}\;
\ForEach(\cmt*[f]{compute layers to delete from $\Ctmp$}){GPU $i \in \mathcal{G}$}{
  $\Mdel(i) \gets \Ctmp[i] \setminus \Ctgt[i]$\;
  $\Bnew(i) \gets$ \MaxBlocks{$i,\, |\Ctgt[i]|$}
}
$\Bnew \gets \min_{i \in \mathcal{G}}\, \Bnew(i)$\;

\SyncCommit{$\Mdel,\, \Bnew$}
\cmt*[f]{commit new config, then delete old layers \& resize KV}\;

\Return $\mathrm{true}$

\caption{\textsc{Async PP Reconfiguration}}

\label{alg:async-reconfig}
% The algorithm transitions from $\Ccur$ to $\Ctgt$ through an \emph{intermediate configuration} $\Ctmp$ in which both old and new layers coexist, enabling continuous inference throughout the migration.}
\end{algorithm}

With these primitives, we present our reconfiguration algorithm. Table~\ref{tab:notation} summarizes the notation used in Algorithm~\ref{alg:async-reconfig}. Algorithm~\ref{alg:async-reconfig} transitions from $\Ccur$ to $\Ctgt$ through an \emph{intermediate configuration} $\mathcal{C}_{\mathrm{tmp}}$ in which each GPU holds the union of its current and target layer sets, allowing inference to continue uninterrupted throughout the migration. The algorithm proceeds in five phases.

\textbf{Phase~1: Feasibility assessment.} For each GPU~$i$, the algorithm constructs an intermediate layer set $\mathcal{C}_{\mathrm{tmp}}[i] = \mathcal{C}_{\mathrm{cur}}[i] \cup \mathcal{C}_{\mathrm{tgt}}[i]$ and computes the maximum feasible KV-block capacity $\Bmax(i)$ under $\mathcal{C}_{\mathrm{tmp}}$. The global KV budget is then conservatively set to the minimum across GPUs, $B_{\mathrm{shrink}}:= \min_{i \in \mathcal{G}} B_{\max}(i)$. If the number of KV blocks currently in use exceeds $B_{\mathrm{shrink}}$, the system cannot accommodate the intermediate PP configuration even after shrinking the KV cache; in this case, the reconfiguration is deemed infeasible and the algorithm aborts.

\textbf{Phase~2: KV resizing.} If $B_{\mathrm{shrink}} < \Bcur$, the intermediate configuration requires reducing KV cache capacity to free memory. The coordinator, therefore, invokes \textsc{CompactKV} to defragment and consolidate live KV blocks, followed by \textsc{ResizeKV} to shrink the KV cache to $B_{\mathrm{shrink}}$ blocks on every GPU, thereby freeing sufficient memory for loading the additional layers in the next phase.

\textbf{Phase~3: Asynchronous weight loading and KV migration.}
The coordinator issues \textsc{AddLayerWeights} to begin loading new layer weights on the destination GPUs. Concurrently, it constructs the migration map $\Mmig$, mapping each source GPU to its destination GPUs and associated layers, and invokes \textsc{StartKVMigration} to begin streaming KV states. Both operations proceed asynchronously alongside ongoing inference.

\textbf{Phase~4: Convergence monitoring.}
The coordinator polls the gap between the last scheduled token index, $T_{\mathrm{sched}}$, and the last synchronized token index, $T_{\mathrm{applied}}[i]$, on each destination GPU. Once this gap falls below the convergence threshold $\tau$ for all GPUs in $\Madd$, the KV state is considered sufficiently up-to-date, and the algorithm proceeds to commit.

\textbf{Phase~5: Commit.}
The coordinator computes the deletion map $\Mdel$ (layers in $\mathcal{C}_{\mathrm{tmp}}$ that are no longer needed under $\Ctgt$) and the post-reconfiguration KV budget $\Bnew$. It then issues \textsc{SyncAndCommit}, which atomically switches the pipeline to $\Ctgt$ and keeps inference running. It subsequently deletes obsolete layer weights and KV cache asynchronously, and resizes the KV cache to $\Bnew$ to reclaim freed memory.

\begin{figure}[]
    \centering
    \includegraphics[width=1\columnwidth]{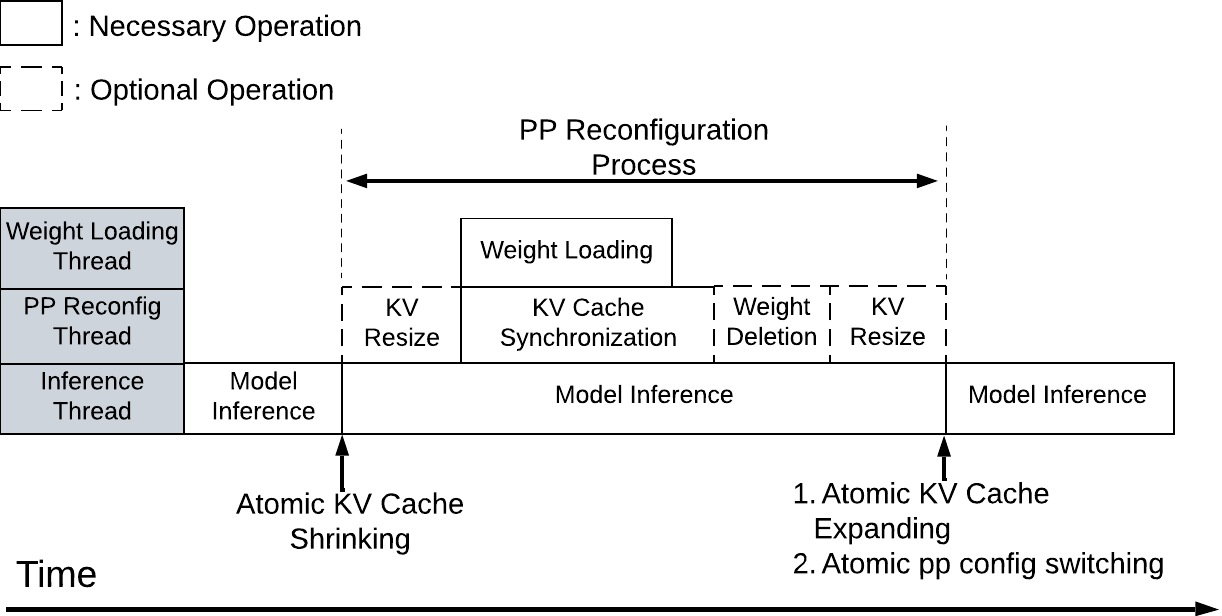}
    \caption{Timeline of asynchronous PP reconfiguration protocol.}
    \label{fig:flexiserve-timeline}
\end{figure}

Figure~\ref{fig:flexiserve-timeline} illustrates the timeline of the live in-place PP reconfiguration protocol across the five phases and highlights the dependency relationships among the primitives. Weight loading and KV cache migration have no direct dependency and can therefore execute in parallel. All other primitives are issued sequentially to ensure correctness and a well-defined transition order.

\section{Worker-Level Dynamic KV Cache Management}
\label{sec:kv_cache_management}
Pipeline reconfiguration requires dynamically adjusting KV cache capacity as layers are reassigned across GPUs. To this end, \textsc{PipeLive} introduces a unified KV cache management design enabling lightweight resizing and leveraging layer stacking to balance PageAttention’s internal fragmentation with layer migration granularity.

\subsection{KV Cache Resizing}

Pipeline reconfiguration changes the number of layers assigned to each GPU, requiring the KV cache to be resized accordingly. In existing systems such as PagedAttention~\cite{kwon2023vllm}, each layer’s KV cache is stored in a preallocated contiguous GPU buffer and accessed at block granularity via a block table. Because GPU memory cannot be resized in place, shrinking or expanding the KV cache requires allocating a new buffer and copying all live KV blocks, making dynamic resizing costly and impractical during reconfiguration.

\begin{figure}[t]
    \centering
    \includegraphics[width=1\columnwidth]{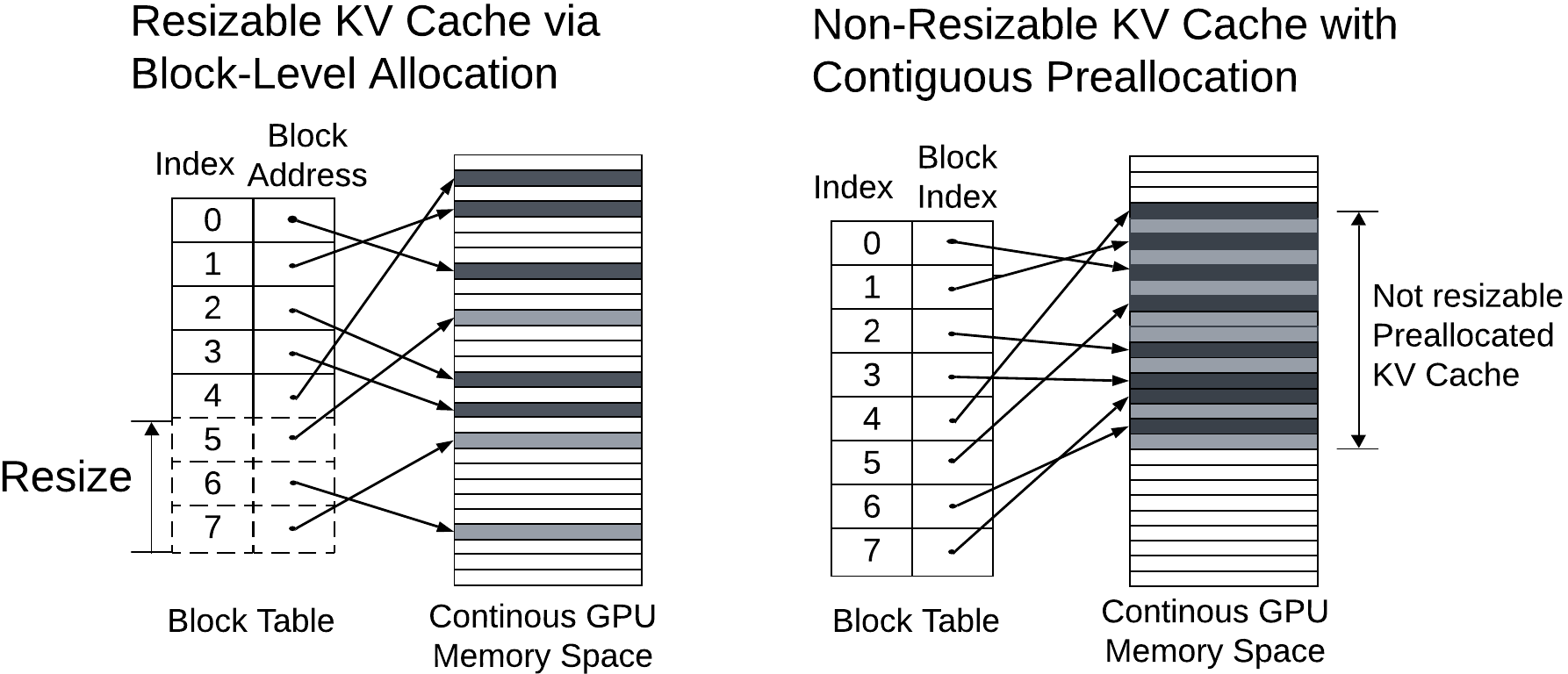}
    \caption{PipeLive extends PagedAttention to access non-contiguous GPU memory, enabling dynamic KV cache resizing during live migration.}
    \label{fig:flexiserve-indirect}
\end{figure}

\textsc{PipeLive} addresses this by adopting a block-level KV cache allocation that decouples logical KV organization from physical memory layout. As shown in Figure~\ref{fig:flexiserve-indirect}, instead of per-layer contiguous buffers, the KV cache is maintained as a list of independently allocated GPU blocks that can be allocated and freed on demand.  For KV cache shrinking, since KV blocks are sparsely allocated, \textsc{PipeLive} performs a compaction process that moves unallocated blocks to the end of the block list and releases them in batches. This process involves only pointer updates and incurs negligible overhead (less than 1ms). For KV cache expansion, \textsc{PipeLive} appends newly allocated KV blocks to the block list.

To support non-contiguous KV access, similar to PageAttention, which uses a block table to map requests to KV block indices, \textsc{PipeLive} stores resolved block addresses directly in the block table. This design enables efficient KV cache resizing during live migration while preserving the access efficiency of PageAttention, with no measurable performance degradation in practice.

\subsection{Layer Stacking}

\textsc{PipeLive} enables KV resizing by allowing PageAttention to index non-contiguous GPU blocks, as discussed in the previous subsection. However, this introduces a key challenge: GPU memory allocation must respect the CUDA virtual-memory allocation granularity~\cite{prabhu2024vattention}, which is commonly 2~MiB on current NVIDIA GPUs\footnote{\href{https://docs.nvidia.com/cuda/cuda-driver-api/group__CUDA__VA.html}{NVIDIA CUDA Driver API, ``Virtual Memory Management''}.}, whereas PageAttention typically uses much smaller KV block sizes (32KB--128KB) to control internal fragmentation. Directly matching KV blocks to the physical allocation unit would therefore lead to severe memory waste due to fragmentation.

\begin{figure}[t]
    \centering
    \includegraphics[width=0.9\columnwidth]{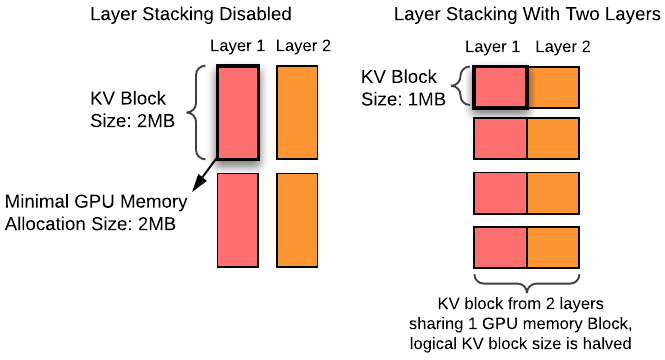}
    \caption{Layer stacking KV cache layout in \textsc{PipeLive}. Stacking two layers into one GPU memory block halved the logical KV block size.}
    \label{fig:layer-stacking}
\end{figure}

To address this, we introduce a \emph{layer stacking} method. Instead of mapping each KV block to a full 2~MiB physical allocation unit, KV blocks from $k$ layers with the same block index are packed into a single physical block (Figure~\ref{fig:layer-stacking}). Let $C$ denote the token capacity of a 2~MiB block under the given model and precision. With stacking factor $k$, each layer effectively occupies $C/k$ tokens within the shared block. This sharing reduces unused space within each allocation unit, improving memory utilization and mitigating internal fragmentation without changing the total KV capacity.

This design introduces a trade-off: increasing $k$ improves memory efficiency but reduces the granularity of PP reconfiguration, as layers must be reconfigured in larger groups. Specifically, layer migration operates at granularity $k$, requiring each partition to be a multiple of $k$. \textsc{PipeLive} selects $k$ to strike a balance between reconfiguration granularity and memory efficiency, as evaluated later in \S\ref{sec:layer_stacking}.

\section{Worker-Level State Synchronization}
\label{sec:state_sync}
A \emph{Reconfiguration Worker} runs on every GPU that performs reconfiguration in model serving. Functionally, each Reconfiguration Worker comprises two components: (1)~a \emph{KV Migrator}, which synchronizes KV cache between source and destination GPUs, and (2)~a \emph{Weight Loader}, which asynchronously loads the weights of newly assigned layers onto the local GPU. We describe each component in turn.

\subsection{KV Migrator}
\label{sec:kv-synchronizer}
During PP reconfiguration, the KV Migrator runs on every GPU and is responsible for receiving KV cache segments migrated to the local device and sending segments to be migrated to other GPUs, while ensuring that migration does not introduce noticeable interference to ongoing inference. To achieve this, we implement concurrent KV transmission and streamed KV synchronization, enabling KV migration to overlap with model inference and proceed in bounded, incremental transfers.

% inter-GPU KV cache transfer is not unique to PP reconfiguration: existing prefill-decode disaggregation (PD Disaggregation) systems~\cite{} use dedicated prefill instances to compute KV states and then transfer them to decode instances. However, kv synchronization has two additional challenges compared to kv transfer in PD Disaggregation.
\textbf{Concurrent KV transmission.} We use NVIDIA Collective Communications Library (NCCL)~\cite{nvidiaNCCL}, a high-performance communication library for GPU-to-GPU data transfer, to migrate KV cache. To avoid interrupting ongoing inference during KV migration, we perform KV cache transfers on a \emph{dedicated NCCL communicator group} that is independent of the communicator group used for pipeline-parallel intermediate-state forwarding. The migration group is further assigned a lower CUDA stream priority so that its GPU work yields to inference kernels. While this design isolates KV transfers from the inference data path, operating two NCCL groups concurrently on overlapping sets of GPUs introduces the risk of \emph{deadlock}. Under NCCL's semantics, where issuing an NCCL operation on one communicator group stalls all other NCCL operations on the same GPU until completion, concurrent operations on separate groups can lead to circular waits if not carefully coordinated.

\begin{figure}[t]
    \centering
    \includegraphics[width=0.8\columnwidth]{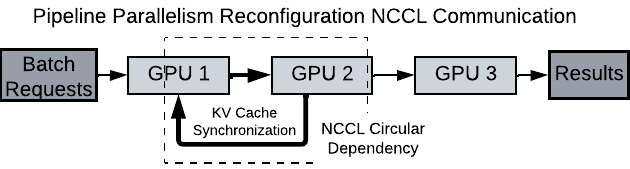}
    \caption{NCCL Communications in PP Reconfiguration. A circular dependency exists between GPU 1 and GPU 2, causing a deadlock.}
    \label{fig:flexiserve-nccl}
\end{figure}

Figure~\ref{fig:flexiserve-nccl} shows the NCCL communication patterns of PP reconfiguration where there are intermediate states transferred between pipeline stages, and GPU~2 migrates the KV cache to GPU~1. This creates a circular wait: for example, GPU~1 blocks on an \texttt{ncclRecv} from the inference group, while GPU~2 blocks on an \texttt{ncclSend} from the migration group targeting GPU~1; neither can make progress. 

To eliminate this deadlock, we introduce a two-phase handshake protocol with asymmetric entry semantics. All NCCL operations on a GPU are serialized via a per-GPU mutex. For intermediate-state transfers between PP stages, operations can proceed immediately after the mutex is acquired, ensuring inference remains prioritized and unblocked. In contrast, KV cache migration follows a two-phase entry protocol. The sender first acquires the mutex and sends an ACK to the receiver via TCP. Upon receiving the ACK, the receiver attempts to acquire its local mutex. If the attempt fails, it responds with REJECT; the sender then releases the mutex and retries after a timeout $\tau$. If successful, the receiver responds with ACCEPT, at which point both parties hold their respective mutexes and can safely initiate the transfer.

This protocol serializes all NCCL operations across both the main inference and KV-migration communicator groups, preventing deadlocks. Since the two-phase handshake is applied only to KV migration, the main inference path incurs negligible overhead.

\textbf{KV Migration with KV Patch} For each GPU pair $(r_s, r_d)$ involved in migration, the KV migrator on each GPU spawns a dedicated \emph{sender thread} on~$r_s$ and a \emph{receiver thread} on~$r_d$. The sender extracts and transmits KV entries for the migrating layers, while the receiver applies incoming entries to the corresponding local layer caches. Multiple sender--receiver pairs may coexist within a single migrator to handle concurrent migrations across different GPU pairs.

Each sender thread maintains a \emph{dirty bitmap} $\mathcal{B}^{(i_s,i_d)} \in {0,1}^{N}$, where $N$ is the total number of physical KV cache slots for the migrating layers, with each slot storing one token. The bitmap is initialized to all zeros. Let $S^{(t)}$ denote the set of physical slot indices written by the inference worker at step~$t$, with $n^{(t)} = |S^{(t)}|$. After each inference step, the inference worker marks the newly written slots as dirty: $\mathcal{B}[s] \gets 1$ for all $s \in S^{(t)}$. The sender thread periodically \emph{drains} the bitmap by atomically extracting the dirty set $\mathcal{D} = \{s: \mathcal{B}[s] = 1\}$, resetting $\mathcal{B}$ to zero, gathering the KV data at the slots in~$\mathcal{D}$ from GPU memory, and transmitting the resulting \emph{KV patch} to~$i_d$. This drain-and-transmit cycle repeats on a best-effort basis until migration completes. On the receiver side, each receiver thread continuously receives incoming patches and writes them into the local layer KV cache.

Concurrently with KV streaming, the Reconfiguration Coordinator performs \emph{convergence tracking} (Phase~4 of Algorithm~\ref{alg:async-reconfig}) to determine when the destination KV cache is sufficiently up-to-date to commit the new PP configuration. The scheduler maintains a cumulative token counter $T_{\mathrm{sched}}$, incremented by $n^{(t)}$ after each inference step. On the receiver side, each thread maintains a counter $T_{\mathrm{applied}}$, incremented upon applying each incoming KV patch. The coordinator periodically queries $T_{\mathrm{applied}}$ for each destination GPU and commits once $T_{\mathrm{sched}} - T_{\mathrm{applied}} < \tau$ holds for all destinations.

In practice, $\tau$ controls the trade-off between service interruption and synchronization overhead. Higher inter-GPU bandwidth enables faster KV synchronization, allowing a smaller $\tau$ to minimize service disruption, while lower bandwidth may require a larger $\tau$. On our testbed, GPUs are connected via InfiniBand with approximately 100 Gbps of bandwidth, and we set $\tau$ to 50 tokens.

\subsection{Weight Loader}
The Weight Loader asynchronously materializes the weights of newly assigned layers during PP reconfiguration. To minimize latency, model weights are preloaded into CPU memory at initialization. Upon migration, the Weight Loader directly transfers weights from CPU to GPU memory, avoiding disk I/O on the critical path, with a fallback to disk-based loading when CPU memory is insufficient.

To prevent interference with ongoing inference, weight loading is executed on a dedicated CUDA stream with lower priority than the inference stream. During PP reconfiguration, the Weight Loader incrementally stages only the parameters needed by the target mapping on each GPU, ensuring timely availability while minimizing contention for GPU resources.

\section{Performance Evaluation}
In this Section, we evaluate \textsc{PipeLive} to quantify the overheads of dynamic pipeline reconfiguration and to isolate the impact of its key design components.

\subsection{Implementation}
We implement \textsc{PipeLive} on top of VLLM and FlashAttention~\cite{dao2022flashattention,dao2023flashattention2}, with approximately 5,000 lines of Python and CUDA/C++ code.

\textbf{Reconfiguration control.}
We implemented \emph{Reconfiguration Coordinator}, which runs outside the vLLM inference loop to orchestrate migration and synchronize GPUs without interfering with the serving path.

\textbf{Migration workers.}
We extend vLLM’s per-GPU worker with a \emph{Reconfiguration Worker} that executes layer migration and KV cache operations. To safely overlap migration with inference, we implement a KV cache migrator with a two-phase handshake that enforces ordering between NCCL groups and avoids deadlocks as discussed in Section~\ref{sec:kv-synchronizer}.

\textbf{Kernel support.}
We extend FlashAttention’s PageAttention to support resolving in-continuous kv block addresses on-the-fly.

\textbf{KV cache layout.}
We replace the default contiguous KV preallocation with block-level allocation to enable KV cache resizing. Furthermore, we introduce a layer-stacked block layout to reduce kv block's internal fragmentation.
\begin{figure*}[t]
    \centering
    \includegraphics[width=0.75\textwidth]{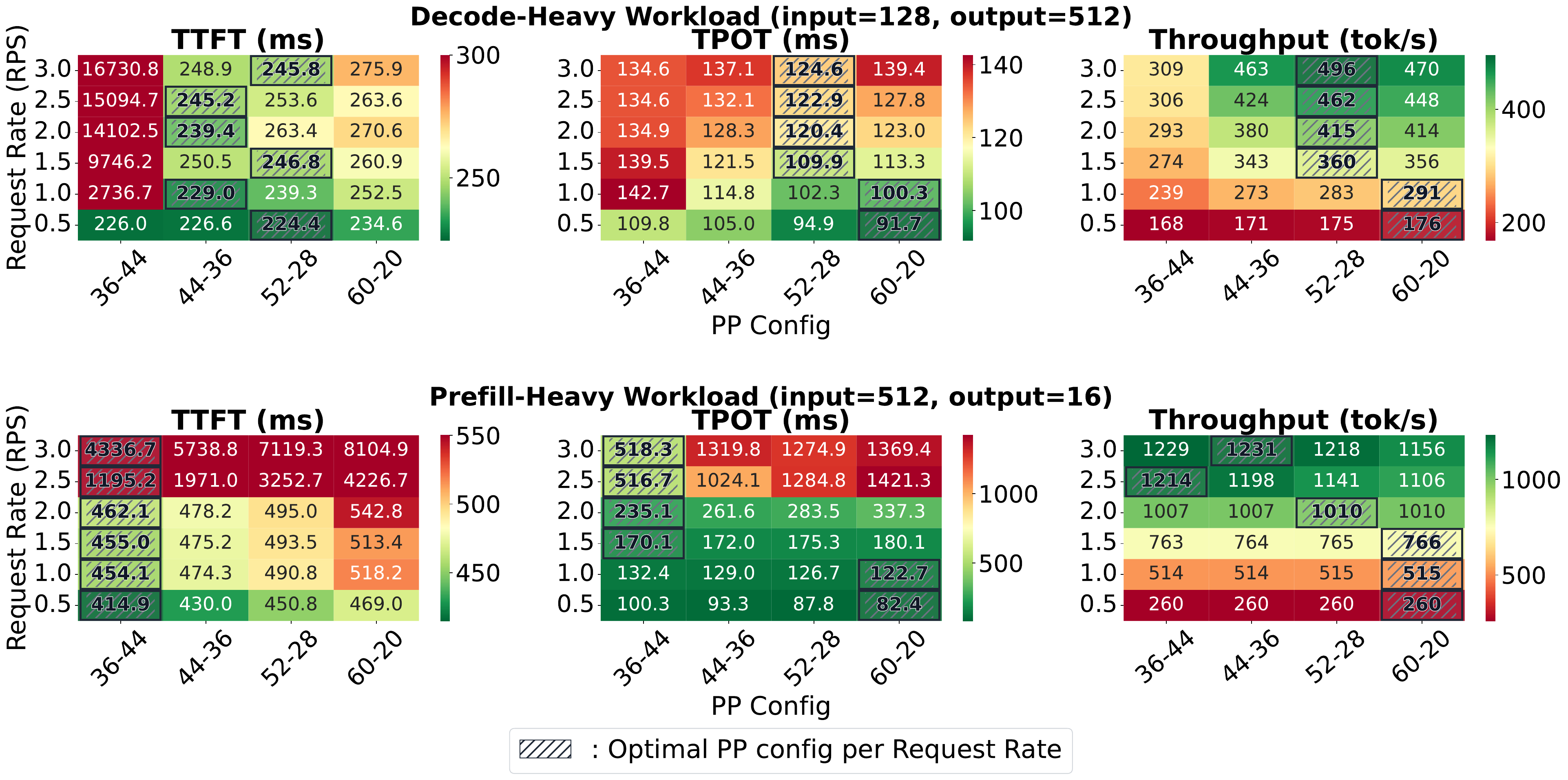}
    \caption{Performance of different PP configurations for heterogeneous GPUs under the PatternShift workload across request rates.}
    \label{fig:heatmap}
\end{figure*}

\subsection{Experimental Setup}
\label{sec:experiment_setup}

To illustrate the efficiency gains enabled by live, in-place PP reconfiguration, we evaluate our approach on a heterogeneous GPU testbed. As discussed in Section~\ref{sec:motivation}, hardware heterogeneity induces workload-dependent optimal PP configurations (e.g., prefill-heavy vs. decode-heavy). This creates natural performance asymmetry across configurations, exposing the benefits of dynamically switching to the workload-aware optimal PP layout.

% \textbf{Testbed.} We construct a heterogeneous GPU testbed consisting of one NVIDIA A100 (80GB) and one NVIDIA L40S (48GB). We show the specs of these two GPUs in figure~\ref{tab:gpu-specs} We deploy the model using a two-stage PP configuration, with the first stage placed on the A100 and the second stage on the L40S. The two GPUs are located on separate nodes and communicate over InfiniBand via NCCL, enabling direct GPU-to-GPU communication across nodes.
{
\textbf{Testbed.} Our primary heterogeneous testbed consists of one NVIDIA A100 (80GB) and one NVIDIA L40S (48GB) located on separate nodes and connected through InfiniBand using NCCL; their specifications are summarized in Table~\ref{tab:gpu-specs}. We deploy the model as a two-stage pipeline, placing the first stage on the A100 and the second on the L40S. The A100's higher memory bandwidth and the L40S's stronger compute capability make the optimal layer partition workload-dependent, providing a representative setting for evaluating live switching between PP configurations optimized for different request mixes.

To evaluate scalability beyond two stages, we use up to 12 A100 GPUs and exercise \textsc{PipeLive}'s ability to expand or shrink the pipeline by adding or removing stages at runtime. This complementary homogeneous setup isolates how migration time and service interruption evolve with pipeline depth, demonstrating whether live reconfiguration remains practical at larger deployment scales.
}

\begin{table}[]
  \centering
  \small
  \caption{Comparison of the two GPUs used in the testbed.}
  \label{tab:gpu-specs}
  \begin{tabular}{lcc}
    \toprule
    \textbf{Metric} & \textbf{A100 80GB} & \textbf{L40S} \\
    \midrule
    Memory bandwidth & 2039 GB/s & 864 GB/s \\
    FP16/BF16 Tensor Core & 624 TFLOPS & 733 TFLOPS \\
    FP8 Tensor Core & N/A & 1466 TFLOPS \\
    GPU memory & 80 GB HBM2e & 48 GB GDDR6 \\
    \bottomrule
  \end{tabular}
\end{table}

\textbf{Metrics and Model.} We evaluate performance using (1) \textit{time-to-first-token (TTFT)}, which captures request latency for initial response generation; (2) \textit{time-per-output-token (TPOT)}, which measures per-token decoding latency; (3) \textit{end-to-end (E2E) latency}; and (4) \textit{fixed-trace completion throughput}. The three latency metrics characterize per-request quality of service (QoS), while throughput captures system-level performance.

To enable holistic comparison across configurations, we introduce a composite \emph{performance score}, denoted by $\mathcal{S}_{\mathrm{perf}}$, over these four metrics. Across all evaluated methods, latency metrics use inverted min--max normalization, $s_x=(x_{\max}-x)/(x_{\max}-x_{\min})$, while throughput uses standard min--max normalization, $s_{TP}=(TP-TP_{\min})/(TP_{\max}-TP_{\min})$. $\mathcal{S}_{\mathrm{perf}}$ gives equal weight to all four components:
\[
\mathcal{S}_{\mathrm{perf}}=
\left(s_{\mathrm{TTFT}}+s_{\mathrm{TPOT}}+s_{\mathrm{E2E}}+s_{TP}\right)/4.
\]
This score provides a unified measure for comparing PP configurations under varying workloads and request rates.

We evaluate end-to-end performance using Llama 3-70B~\cite{grattafiori2024llama3}, a widely adopted open-source model whose scale is sufficient to expose meaningful system bottlenecks while remaining practical to serve on our testbed.
% Previous figure retained for reference.
\iffalse
\begin{figure*}[t]
    \centering
    \includegraphics[width=1\textwidth]{Figures/e2e_llama_qwen_combined.pdf}
    \caption{Performance of different PP configurations for heterogeneous GPUs under the PatternShift workload.}
    \label{fig:e2e-config-cmp}
\end{figure*}
\fi

{
\textbf{Baselines.} Prior work supports only limited forms of PP reconfiguration, and no existing open-source LLM inference system provides in-place, fine-grained PP reconfiguration, leaving no directly comparable dynamic baseline. In the end-to-end experiment, we therefore compare \textsc{PipeLive} against all feasible static PP configurations to determine whether live reconfiguration improves overall serving performance. We additionally evaluate variants of \textsc{PipeLive} with selected components disabled to isolate the contribution of each design feature.
}

% In the end-to-end experiment, We therefore construct strong static baselines by selecting representative fixed PP configurations that are optimal under different workload regimes.

% Specifically, we consider three static configurations: (1) \emph{Prefill-Optimal}: the optimal configuration under prefill-heavy workloads, (2) \emph{Decode-Optimal}: the optimal configuration under decode-heavy workloads, and (3) \emph{Balanced}: a globally optimal configuration that balances performance across workloads.

{
% \textbf{Workloads.} For end-to-end evaluation, we construct a 15-minute workload trace from the raw 24-hour BurstGPT trace. We uniformly sample requests and preserve their original temporal order. This trace retains BurstGPT's time-varying arrival pattern and uses a single global scaling factor to match the load intensity to our heterogeneous A100--L40S testbed. We calibrate this factor so that the balanced configuration, which distributes the model layers evenly across the two GPUs, remains around 60\% peak KV-cache utilization. The resulting 1,350-request workload preserves the trace's natural load and request-mix shifts while creating enough memory pressure to expose configuration-dependent behavior.在进行实验的时候,我们将会对所有的static configuration进行profile,然后在每一段的rolling window里面选出最好的配置,并且让pipeline在这些最佳的配置里面这些window里面执行最佳配置的切换。 这能够帮助我们测试pipeline的efficiency, pipeline是否足够的efficient来支持在optimal configuration进行切换, 然后并且取得对应的performance。
\textbf{Workloads.} Two workloads are used for the end-to-end experiment and the performance ablation evaluation of \textsc{PipeLive}. The end-to-end evaluation uses a 15-minute, 1,350-request workload sampled from the raw 24-hour BurstGPT trace. The sampled trace preserves BurstGPT's time-varying arrival pattern, with its load intensity scaled to the testbed. The scaling factor is calibrated such that the balanced configuration, which evenly distributes model layers across GPUs, reaches approximately 60\% peak KV-cache utilization. This workload retains the trace's natural load dynamics while creating sufficient memory pressure. Building on this setup, profiling all feasible static PP configurations over the trace reveals window-level changes in the workload-preferred configuration, and \textsc{PipeLive} performs reconfiguration based on these changes, enabling an end-to-end evaluation of whether switching among workload-preferred configurations improves performance. The configuration-selection procedure and resulting reconfiguration points are described in the end-to-end evaluation.

BurstGPT captures realistic workload dynamics, but the co-variation in request rate, arrival burstiness, and token lengths makes individual transitions difficult to isolate. For the ablation evaluation, we therefore introduce the \textbf{PatternShift workload}, a two-phase benchmark with a configurable request rate. Each 200-request run transitions at its midpoint from a prefill-heavy phase with 512/16 input/output tokens to a decode-heavy phase with 128/512 tokens. For each phase and request rate, we profile the feasible static PP configurations and configure \textsc{PipeLive} to switch between the phase-optimal configurations at this known boundary. This deterministic setup provides a common time origin for measuring reconfiguration overhead and transient performance effects across different request rates. Figure~\ref{fig:heatmap} reports performance across different request rates for both the prefill-heavy and decode-heavy phases of the PatternShift workload, using a four-layer partition granularity. The results show that the optimal partition shifts substantially with both workload pattern and request rate, while different metrics may favor different partitions.

}

% Enabling live, in-place PP reconfiguration would allow the system to dynamically switch between configurations as workload conditions change, thereby improving throughput and latency metrics (including TTFT and TPOT).

% During execution, we trigger \textsc{PipeLive} to reconfigure between optimal PP configurations at the transition point, where the target configuration is determined by the performance score defined in \S\ref{sec:experiment_setup}.

% Optimal PP configuration is inherently dynamic, shifting with both workload characteristics and request rate. Decode-heavy workloads favor allocating more layers to the A100 due to its higher memory bandwidth, while prefill-heavy workloads, especially at higher request rates, benefit from assigning more layers to the L40S, which offer stronger compute capability. Therefore 

\begin{table*}[t]
  \centering
  \setlength{\tabcolsep}{2.5pt}
  \renewcommand{\arraystretch}{1.12}
  \small
  \resizebox{\textwidth}{!}{%
  \begin{tabular}{l@{\hspace{5pt}}rrr@{\hspace{5pt}}rrr@{\hspace{5pt}}rrr@{\hspace{5pt}}rr}
    \toprule
    & \multicolumn{3}{c}{TPOT (ms/token) $\downarrow$}
    & \multicolumn{3}{c}{TTFT (ms) $\downarrow$}
    & \multicolumn{3}{c}{E2E Latency (ms) $\downarrow$}
    & \multicolumn{1}{c}{Throughput}
    & $\mathcal{S}_{\mathrm{perf}}$ $\uparrow$ \\
    \cmidrule(lr){2-4}\cmidrule(lr){5-7}\cmidrule(lr){8-10}
    Configuration & Mean & P50 & P99
    & Mean & P50 & P99 & Mean & P50 & P99
    & (tok/s) $\uparrow$ & \\
    \midrule
    Static 36/44
      & 212.47 & 170.79 & \textbf{1,111.08}
      & 1,970.50 & 362.89 & 20,712.94
      & 7,134.47 & 2,393.34 & 45,796.83 & 485.03 & 0.261 \\
    Static 40/40
      & 234.45 & 172.72 & 1,485.60
      & \textbf{585.84} & \textbf{360.34} & 3,890.67
      & 6,096.46 & 2,065.92 & 49,985.09 & 484.86 & 0.533 \\
    Static 44/36
      & 243.23 & 158.15 & 1,687.60
      & 618.60 & 368.87 & 3,966.66
      & 5,832.16 & 2,098.07 & 49,097.05 & 485.47 & 0.571 \\
    \textbf{Static 48/32}
      & 248.19 & 149.38 & 1,763.19
      & 665.63 & 376.86 & 4,437.27
      & 5,742.71 & 2,176.93 & \textbf{45,217.42} & 486.90 & 0.668 \\
    Static 52/28
      & 265.84 & 142.68 & 2,092.24
      & 729.79 & 376.14 & 4,987.69
      & 5,710.59 & 2,170.60 & 45,278.52 & 487.62 & 0.638 \\
    Static 56/24
      & 279.98 & 138.19 & 2,127.10
      & 794.93 & 400.59 & 5,517.72
      & 5,793.55 & 2,266.41 & 44,706.34 & 487.93 & 0.698 \\
    Static 60/20
      & 302.22 & 146.72 & 2,337.65
      & 837.75 & 402.22 & 5,841.98
      & 6,112.64 & 2,391.48 & 46,804.68 & 488.15 & 0.586 \\
    \midrule
    \textsc{PipeLive}
      & \textbf{211.82} & \textbf{140.61} & 1,170.10
      & 598.83 & 363.53 & \textbf{3,420.80}
      & \textbf{5,245.97} & \textbf{2,001.75} & 45,684.82
      & \textbf{487.92} & \textbf{0.998} \\
    Gain over the best static conf.
      & $+14.65\%$ & $+5.88\%$ & $+33.64\%$
      & $+10.04\%$ & $+3.54\%$ & $+22.91\%$
      & $+8.65\%$ & $+8.05\%$ & $-1.03\%$
      & $+0.21\%$ & $+49.39\%$ \\
    \bottomrule
  \end{tabular}%
  }
  \caption{End-to-end performance on the BurstGPT
  workload. PP configurations denote the
  A100/L40S layer split. Bold metric values are best in each column; the
  bold configuration, Static 48/32, is the highest-scoring static. ``Gain over
  the best static conf.'' reports \textsc{PipeLive}'s improvement over Static 48/32, where positive values
  denote lower latency, higher throughput, or a higher
  $\mathcal{S}_{\mathrm{perf}}$.}
  \label{tab:day20-unique1350-kvcached}
  \vspace{2pt}
  % \begin{minipage}{0.99\textwidth}
  %   \footnotesize
  %   \textit{Notes.} Configurations report the A100/L40S layer split. \textsc{PipeLive}
  %   follows the reconfiguration points selected by the 60-second profiling
  %   policy.
  %   TTFT, TPOT, and E2E use inverted min--max normalization, while
  %   completion throughput uses standard min--max normalization across all
  %   six evaluated methods. $\mathcal{S}_{\mathrm{perf}}=(s_{\mathrm{TTFT}}+
  %   s_{\mathrm{TPOT}}+s_{\mathrm{E2E}}+s_{TP})/4$. Static 48/32 is the
  %   highest-scoring static.
  %   Bold values are best in each metric column. Positive gain denotes lower
  %   latency, higher throughput, or a higher score. The superseded sampled Prometheus throughput is
  %   not used. Results are from one repetition per configuration.
  % \end{minipage}
\end{table*}

\subsection{Evaluation of End-to-end Performance}
\begin{figure*}[t]
    \centering
    \includegraphics[width=0.8\textwidth]{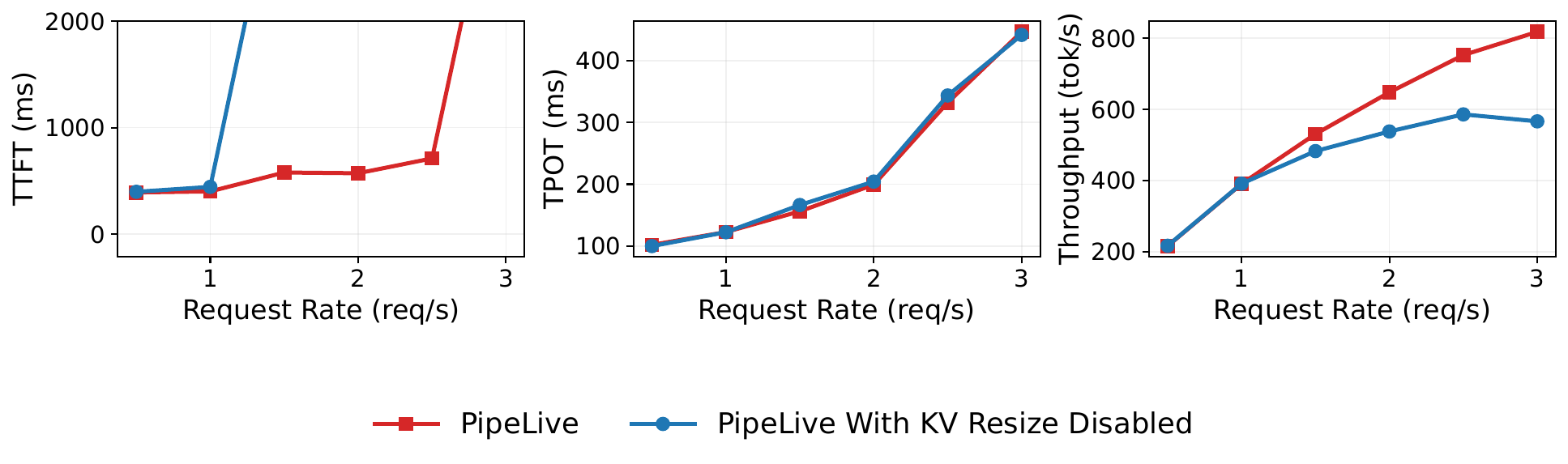}
    \caption{End-to-end performance of PP reconfiguration with KV resizing disabled and enabled under the PatternShift workload.}
    \label{fig:kv-resize}
\end{figure*}

{
This experiment evaluates whether \textsc{PipeLive} can improve end-to-end performance on a real-world workload trace by switching among perferred configurations in different workload windows. Table~\ref{tab:day20-unique1350-kvcached} compares \textsc{PipeLive} against seven feasible static PP configurations on the BurstGPT workload. While each static baseline runs with a fixed configuration throughout the experiment, \textsc{PipeLive} performs reconfiguration at selected points to switch between different static PP configurations.

\textbf{Reconfiguration policy.} Reconfiguration points are derived using a lightweight, profile-guided rule. Each feasible static PP configuration is first profiled independently on the testbed. The resulting profiles are partitioned into aligned 60-second windows, which are long enough to provide stable aggregate estimates while remaining short enough to capture minute-scale workload shifts. Within each window, $\mathcal{S}_{\mathrm{perf}}$ is computed for every configuration, and the one with the highest score is selected. A reconfiguration point is introduced whenever the selected configuration changes between adjacent windows. This procedure identifies three reconfiguration points. Please note that this work focuses on the system mechanisms required to execute live PP reconfiguration, as well as their runtime performance and overhead. The profile-guided policy provides representative target configurations for this evaluation; an online policy that automatically detects workload changes and selects target configurations is complementary and left to future work.

\textbf{Results.} \textsc{PipeLive} achieves the best value in six of the ten reported metrics: mean and P50 TPOT, P99 TTFT, mean and P50 E2E latency, and completion throughput. Static 48/32 is the strongest aggregate static baseline, with $\mathcal{S}*{\mathrm{perf}}=0.668$. Relative to this baseline, \textsc{PipeLive} improves all mean and median latency metrics by 3.54--14.65\%, reduces P99 TTFT and TPOT by 22.91\% and 33.64\%, respectively, and maintains the highest throughput. Collectively, these improvements increase $\mathcal{S}*{\mathrm{perf}}$ from 0.668 to 0.998, a 49.39\% gain, showing that live PP reconfiguration enables \textsc{PipeLive} to capture phase-specific performance benefits at a reconfiguration cost substantially smaller than the resulting gains.
}

\subsection{Effects of KV Resizing}

Different PP configurations induce substantially different KV capacity on each GPU due to changes in layer placement. As a result, a configuration that is feasible under one workload phase may exceed KV capacity after reconfiguration under another phase. Without resizing, this mismatch leads to KV overloading and severe performance degradation.

Figure~\ref{fig:kv-resize} compares PP reconfiguration with and without KV resizing under the PatternShift workload across varying request rates. Without resizing, the KV allocation fixed to the source configuration becomes insufficient after the workload shifts, causing KV overloading even at modest load (request rate > 1) and resulting in sharp TTFT degradation.

With KV resizing enabled, the system dynamically adjusts KV capacity during migration to match the target configuration. This eliminates overloading, stabilizes TTFT up to a request rate of 2.5, and significantly improves throughput over 45\% under high load. These results indicate that KV resizing enables reconfiguration to remain both feasible and effective under dynamic workloads by aligning KV capacity with each configuration's requirements.

\subsection{Effects of Layer Stacking to Reduce GPU Memory Fragmentation}
\label{sec:layer_stacking}

To support KV cache resizing, we allocate KV memory at block granularity per layer rather than as a single contiguous GPU buffer. Because GPU allocation has a minimum granularity (e.g., 2MB) while each layer’s KV cache is much smaller, this leads to substantial internal fragmentation.

\begin{figure}[]
    \centering
    \includegraphics[width=0.6\columnwidth]{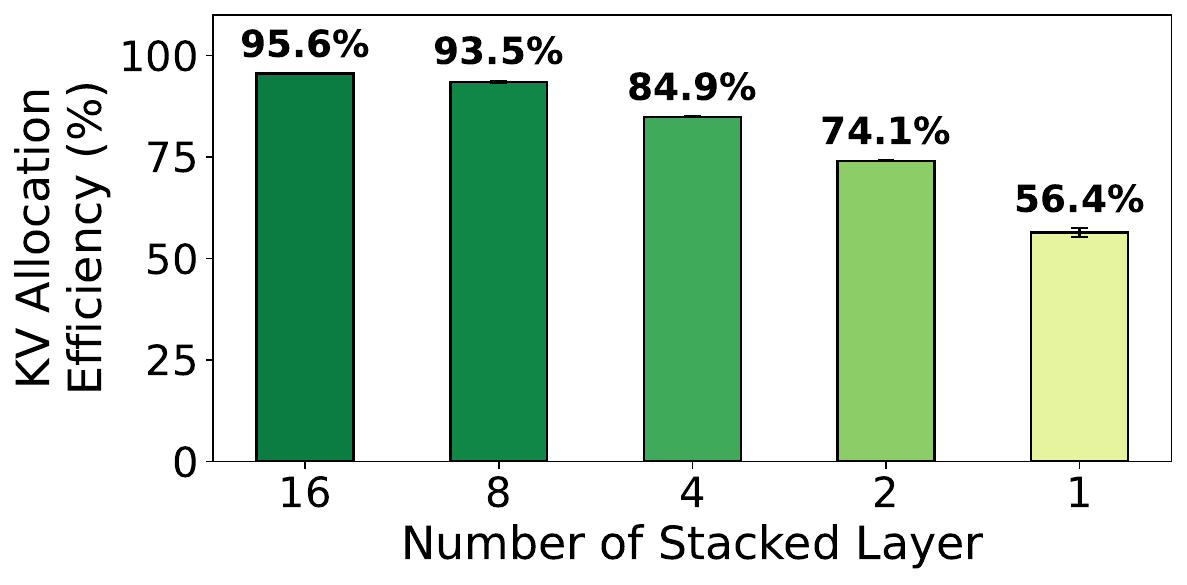}
    \caption{Effective KV utilization with different numbers of stacked layers.}
    \label{fig:layer-stacking-kv-util}
\end{figure}

We address this with \emph{layer stacking}, which packs the KV caches from multiple layers into a single physical GPU block, thereby reducing fragmentation. Figure~\ref{fig:layer-stacking-kv-util} reports \emph{Effective KV Utilization}, defined as the fraction of request-allocated KV cache that is actually consumed by tokens. This metric captures the impact of internal fragmentation. Without stacking, KV utilization reaches only 56\%, indicating that nearly half of the KV memory is lost to fragmentation. Increasing the stacking factor reduces the effective KV block size and correspondingly lowers fragmentation.

%This improvement, however, comes at a cost. A larger stacking factor coarsens the granularity of PP reconfiguration, since stacked layers must be migrated together. This restricts the system’s ability to reach the optimal configuration under changing workloads.

\begin{figure}[]
    \centering
    \includegraphics[width=0.6\columnwidth]{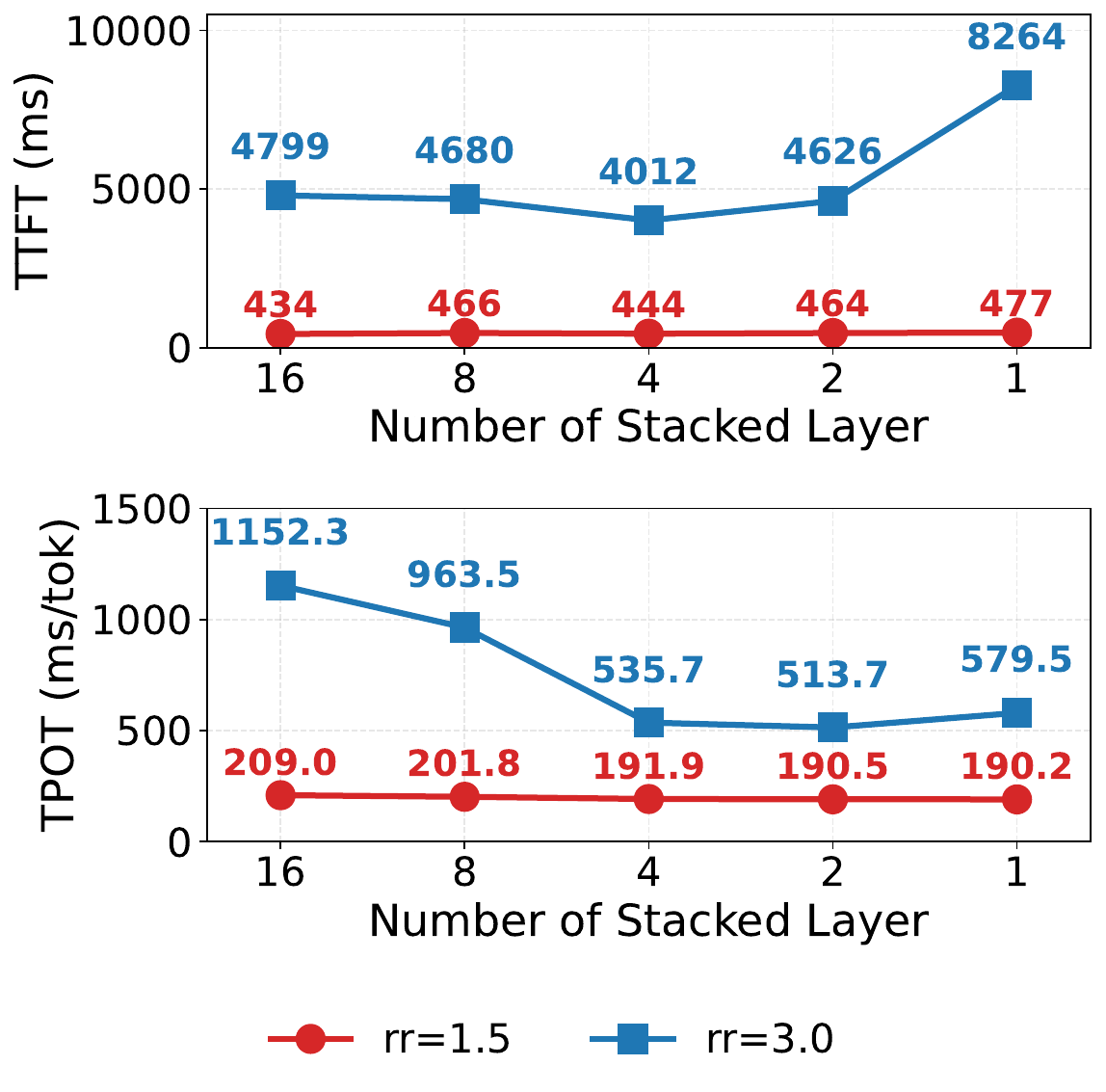}
    \caption{End-to-end performance of PP reconfiguration with different numbers of stacked layers. When it equals 1, layer stacking is disabled.}
    \label{fig:layer-stacking-granularity-cmp}
\end{figure}

\begin{figure}[]
    \centering
    \includegraphics[width=1\columnwidth]{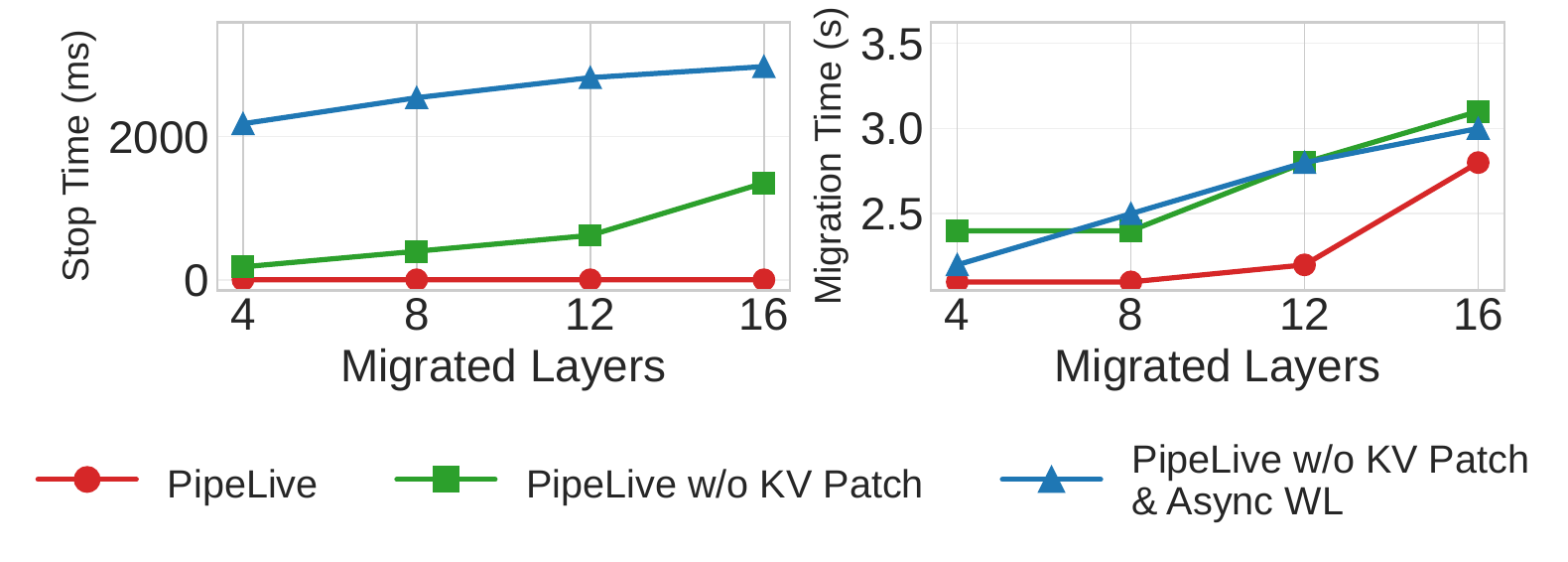}
    \caption{Stop time and migration time as the number of migrated layers increases. }
    \label{fig:flexiserve-migration-layer-stop-time}
\end{figure}
\begin{figure}[]
    \centering
    \includegraphics[width=0.9\columnwidth]{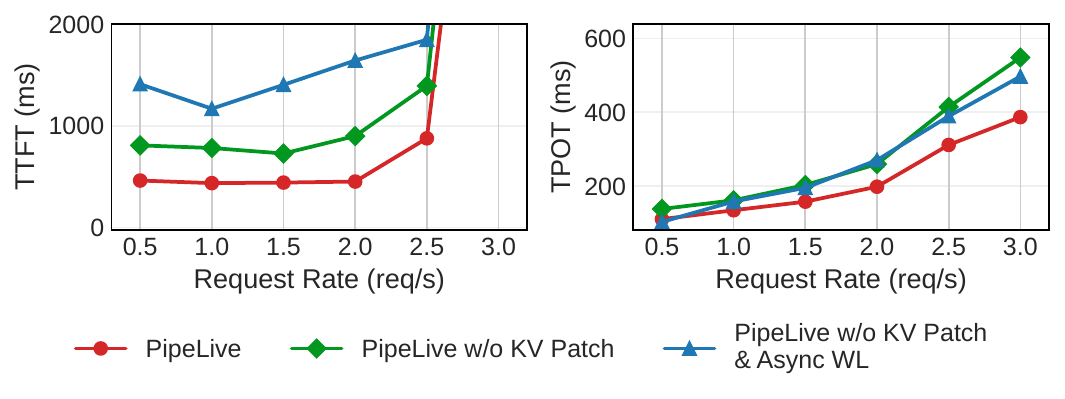}
    \caption{End-to-end tests of performance of PP reconfiguration with different migration modes.}
    \label{fig:flexiserve-migration-mode}
\end{figure}
This improvement, however, comes at a cost. Figure~\ref{fig:layer-stacking-granularity-cmp} quantifies this trade-off on LLaMA-70B under the PatternShift workload at request rates 1.5 and 3.0. When stacking is disabled, high KV cache wastage directly translates into performance degradation under heavy load; for example, TTFT increases by 51\% compared to a stacking factor of 4. In contrast, overly large stacking factors limit reconfiguration flexibility, preventing the system from adapting to the optimal configuration and increasing both TTFT and TPOT.

These results reveal a clear trade-off: insufficient stacking wastes memory, while excessive stacking reduces adaptability. A moderate stacking factor achieves both low fragmentation and sufficient reconfiguration flexibility. In practice, we select a stacking factor of 4 as the default in \textsc{PipeLive}.

\subsection{Effects of Asynchronous Weight Loading and KV Patch Mechanism}

\textsc{PipeLive} employs asynchronous weight loading with a KV patch mechanism to enable non-blocking migration. During reconfiguration, weights are loaded in the background while KV states are continuously synchronized, minimizing service interruption and maintaining inference progress.

We evaluate three settings under the PatternShift workload: full \textsc{PipeLive}, \emph{PipeLive w/o KV Patch}, which retains asynchronous weight loading, and \emph{PipeLive w/o KV Patch \& Async WL}, which disables KV patching and inference-overlapped asynchronous weight loading. Figure~\ref{fig:flexiserve-migration-layer-stop-time} how migration time, i.e., the interval from the initiation of reconfiguration until the new PP configuration is fully in effect, and stop time, i.e., the interval for which inference is interrupted vary with the number of migrated layers under the PatternShift workload. Across all migration sizes, \textsc{PipeLive} keeps stop time below 1.5ms, while PipeLive w/o KV Patch stalls for 181--1,353ms and PipeLive w/o KV Patch \& Async WL stalls for 2,181--2,978ms. Total migration time remains between 2.1s and 3.1s across all three modes.

Figure~\ref{fig:flexiserve-migration-mode} further evaluates performance within a $\pm$15s window around migration, which covers \textsc{PipeLive}'s migration process. Enabling KV patch alone improves TTFT by up to 49.7\% and TPOT by up to 29.5\%. Enabling both asynchronous loading and KV patch further improves TTFT by up to 72.4\%, while achieving up to 26.7\% improvement in TPOT.

Overall, \textsc{PipeLive} converts migration from a blocking operation into a largely background activity, reducing service interruption from seconds to below 1.5ms while preserving end-to-end performance even when the system is handling substantial request load.

% Also we have done a test to showcase the performance of different migration mode with different migration frequency. The results show that with the KV patch mechanism, the stop time can be significantly reduced especially when the migration frequency is high(e.g., 1 migration per 100 requests), which leads to better performance during PP reconfiguration.

% Figure~\ref{fig:flexiserve-migration-layer-stop-time} breaks down the stop time and migration time of each migration mode. The results show that with the KV patch mechanism, the stop time can be significantly reduced, which leads to better performance during PP reconfiguration.

\subsection{Scalability Test}
\begin{figure}[]
    \centering
    \includegraphics[width=0.9\columnwidth]{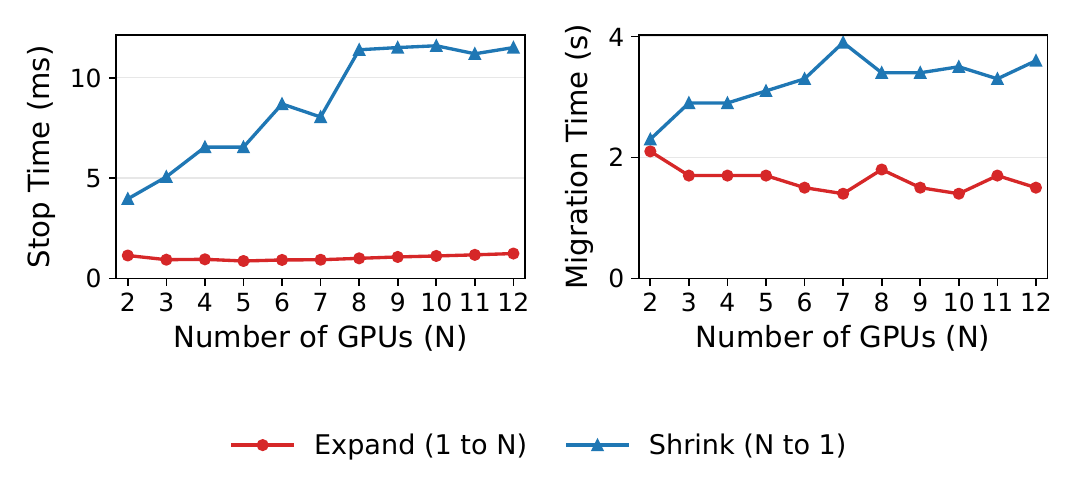}
    \caption{Stop time and migration time when expanding from one GPU to $N$ GPUs and shrinking from $N$ GPUs to one under the PatternShift workload.}
    \Description{Two line charts compare expansion and shrink from two to twelve GPUs. Expansion stop time stays near one millisecond, shrink stop time reaches about twelve milliseconds, and all migration times remain below four seconds.}
    \label{fig:autoscaling-reconfiguration-times}
\end{figure}
{
% The end-to-end evaluation and mechanism microbenchmarks use a two-GPU testbed. The placement space grows combinatorially with the number of stages: at two-layer granularity, moving from two to four GPUs increases the number of candidate placements from tens to hundreds. Selecting a target efficiently from this larger space is a separate optimization problem. We therefore use a separate scalability experiment to demonstrate how \textsc{PipeLive}'s reconfiguration cost changes with the number of participating GPUs.

Beyond redistributing layers across a fixed set of GPUs, \textsc{PipeLive} can add or remove PP stages to vertically scale a model instance. We exercise this capability on up to 12 $\times$ A100 GPUs to evaluate \textsc{PipeLive}'s performance at a larger system scale, using the same BurstGPT-derived PatternShift workload at a moderately high request rate of 3. At the workload midpoint, the instance either expands from one GPU to a near-even $N$-GPU layer placement or shrinks from that placement back to one GPU, where$N \in \{2,\ldots,12\}$. This controlled setup exposes the performance and overhead of live PP reconfiguration as the pipeline depth increases.

Figure~\ref{fig:autoscaling-reconfiguration-times} reports total migration time and stop time. Expansion completes in 1.4--2.1\,s and remains nearly flat as $N$ increases. This trend is consistent with parallelizing weight loading and KV-cache transfer across more destination GPUs, thereby exploiting greater aggregate bandwidth. Shrinking takes 2.3--3.9\,s because model weights and live KV state from multiple source ranks must converge onto one destination GPU, but its cost does not grow proportionally with $N$. Although shrink stop time increases as more stages are retired, it plateaus near 11.5\,ms beyond eight GPUs. Across all scales, stop time remains 0.87--1.24\,ms for expansion and at most 11.59\,ms for shrinking. Thus, every reconfiguration completes within 3.9\,s while exposing inference to only millisecond-scale interruption. Overall, \textsc{PipeLive} preserves efficient, low-disruption in-place reconfiguration as the pipeline scales, making frequent adaptation practical on larger GPU deployments.
}

\section{Related Work}
\textbf{General LLM serving systems.}
Early systems such as Orca established iteration-level scheduling and continuous batching for transformer serving~\cite{yu2022orca,vaswani2017attention}. Subsequent systems each targeted a key serving bottleneck: Sarathi-Serve co-schedules prefills and decodes via chunked prefills~\cite{agrawal2024sarathiserve}; vLLM improves memory efficiency with PagedAttention and fine-grained KV-cache management~\cite{kwon2023vllm}; SGLang reuses KV states through prefix caching with RadixAttention~\cite{zheng2024sglang}; DistServe separates prefill and decode paths to improve goodput~\cite{zhong2024distserve}; and Llumnix focuses on elastic scheduling and instance management~\cite{sun2024llumnix}. Other work studies hardware-aware and heterogeneous serving optimization~\cite{griggs2024melange,he2025uellm,moHetis2025,wagenlander2024tenplex}. Collectively, these systems improve scheduling, memory efficiency, and deployment awareness within fixed execution layouts. \textsc{PipeLive} is complementary: it enables changing the pipeline-parallel layout live and in place while requests remain in flight.

\noindent\textbf{Pipeline parallelism.}
Pipeline parallelism is a standard mechanism for executing models that exceed a single device's memory capacity by partitioning layers across GPUs and forwarding activations stage-by-stage. Foundational systems such as GPipe, Megatron-LM, PipeDream, and Alpa have explored how to partition models, schedule execution across stages, and combine PP with other forms of parallelism~\cite{huang2019gpipe,shoeybi2020megatronlm,narayanan2019pipedream,zheng2023alpa}. This line of work established the system's substrate for large-model distributed execution, but it mostly treats the PP partition as an offline choice made before execution begins. This assumption is increasingly limiting for LLM serving. In practice, the best PP layout depends on the workload mix, latency objectives, request rate, hardware heterogeneity, and KV-cache pressure, all of which may vary over time. \textsc{PipeLive} builds directly on the PP abstraction, but departs from prior work by treating the partition as a dynamic runtime object rather than a static deployment-time decision.

\noindent\textbf{Pipeline-parallel reconfiguration.}
The closest prior efforts that attempt to adapt PP configurations at runtime are HydraServe, FlexPipe, and DynaPipe~\cite{lou2025hydraserve,lin2025flexpipe,xudynapipe}. DynaPipe only supports small-scale redistribution of model layers in the tail pipeline stages, without accounting for the substantial overhead of KV migration or GPU memory constraints during layer redistribution, and therefore cannot support in-place switching between PP configurations with large differences. HydraServe and FlexPipe adjust the set of GPUs participating in pipeline parallelism, but also do not consider GPU memory constraints and do not support in-place PP reconfiguration that redistributes model layers across an existing set of GPUs.

In contrast, \textsc{PipeLive} enables general, live, in-place PP reconfiguration by redistributing model layers and KV cache across existing GPUs via a reconfiguration protocol with dynamic KV resizing, without interrupting ongoing inference. In heterogeneous GPU environments, this capability enables switching between workload-optimal PP configurations that exhibit significant differences in layer distribution.

Prior approaches assume static per-layer KV cache allocation, leading to underutilized memory and limited scalability. Systems such as KVcached~\cite{xing2025towards} enable dynamic KV cache allocation via NVIDIA's virtual memory management (VMM), but require remapping virtual to physical addresses at each inference step, incurring runtime overhead. Moreover, they do not align KV block sizes with the GPU allocation unit, making them unsuitable for efficient KV cache reclamation during in-place PP reconfiguration.

In contrast, \textsc{PipeLive} enables PageAttention to access non-contiguous memory without remapping, and leverages layer stacking to align KV block sizes with GPU allocation unit, reducing internal fragmentation and enabling instantaneous KV block reclamation during resizing. Moreover, it employs KV patching to continuously synchronize KV states, minimizing divergence during migration.

\section{Conclusions and Future Directions}

In this paper, we presented \textsc{PipeLive}, a system that enables efficient live, in-place PP reconfiguration for LLM serving. By dynamically switching between optimal PP configurations, \textsc{PipeLive} reduces mean TTFT, TPOT, and end-to-end latency by 10.04\%, 14.65\%, and 8.65\%, respectively, while reducing P99 TTFT and TPOT by 22.91\% and 33.64\% in heterogeneous GPU deployments. Through dynamic KV resizing and layer stacking, \textsc{PipeLive} sustains up to 2.5$\times$ higher request rates without KV overloading. In addition, inspired by live VM migration, we introduce a KV patching mechanism that reduces TTFT by up to 49.7\% and TPOT by up to 29.5\% during reconfiguration, while keeping service interruption to around 10ms. Together, these techniques make live, in-place PP reconfiguration practical for dynamically adapting to optimal configurations under changing workloads in heterogeneous GPU environments, and lay the foundation for broader scenarios such as multi-tenant LLM serving under dynamic workloads.

While this work focuses on the efficient execution of live, in-place PP reconfiguration, exploring higher-level decisions, such as selecting optimal PP configurations, remains an important direction for future work. We plan to develop reconfiguration algorithms for \textsc{PipeLive} to dynamically optimize PP configurations based on workload dynamics, request rates, available computational resources, and evolving user QoS requirements. Future work could also explore jointly optimizing PP configurations with other forms of parallelism, such as tensor and data parallelism, to further improve resource utilization and performance across diverse deployment settings.
% \begin{figure}[]
%     \centering
%     \includegraphics[width=0.8\columnwidth]{Figures/eval-migration-mode-nreq-true-norm-latency.pdf}
%     \caption{End to end tests of performance of PP reconfiguration with differenrt migration mode and migration frequency.}
%     \label{fig:flexiserve-migration-mode}
% \end{figure}

% \subsection*{Resizable Page Attention}
% In this section, we compare the end-to-end performance of PP reconfiguration with PipeLive's block-level pre-allocation and direct block-table addressing against a baseline implementation that uses the original PageAttention kernel with contiguous preallocation. Both implementations are evaluated under the PatternShift workload and the same PP configuration change. The results in Figure~\ref{fig:flexiserve-pageattention} show that PipeLive's design achieves comparable or even better performance to the original PageAttention kernel while enabling efficient KV cache resizing during PP reconfiguration, demonstrating the effectiveness of our approach in supporting dynamic memory management without sacrificing inference throughput.
\subsection*{}
% \begin{figure*}[]
%     \centering
%     \includegraphics[width=1\textwidth]{Figures/eval-kernel-cmp.pdf}
%     \caption{End-to-end PP Reconfiguration Performance with Flexiserve and PageAttention Kernal}
%     \label{fig:flexiserve-pageattention}
% \end{figure*}
% A subtlety arises because the bitmap deduplicates overlapping slot writes: $|\mathcal{D}| \le \sum_t n^{(t)}$. If the receiver were to increment $T_{\mathrm{applied}}$ by $|\mathcal{D}|$, it would under-count relative to $T_{\mathrm{sched}}$, causing convergence to stall. To keep the two counters aligned, each bitmap update also accumulates $n^{(t)}$ into a separate counter $n_{\mathrm{acc}}$. When the sender drains the bitmap, it attaches $n_{\mathrm{acc}}$ to the outgoing patch and resets the accumulator. The receiver then increments $T_{\mathrm{applied}}$ by $n_{\mathrm{acc}}$ rather than $|\mathcal{D}|$, ensuring faithful progress tracking.

\balance
\bibliographystyle{ACM-Reference-Format}
\bibliography{bibliography}

@inproceedings{kwon2023vllm,
  title = {Efficient Memory Management for Large Language Model Serving with {PagedAttention}},
  author = {Kwon, Woosuk and Li, Zhuohan and Zhuang, Siyuan and Sheng, Ying and Zheng, Lianmin and Yu, Cody Hao and Gonzalez, Joseph and Zhang, Hao and Stoica, Ion},
  booktitle = {Proceedings of the 29th Symposium on Operating Systems Principles (SOSP '23)},
  year = {2023},
}

@article{shoeybi2020megatronlm,
  title = {Megatron-{LM}: Training Multi-Billion Parameter Language Models Using Model Parallelism},
  author = {Shoeybi, Mohammad and Patwary, Mostofa and Puri, Raul and LeGresley, Patrick and Casper, Jared and Catanzaro, Bryan},
  journal = {arXiv preprint arXiv:1909.08053},
  year = {2020},
}

@article{xing2025towards,
  title={Towards Efficient and Practical GPU Multitasking in the Era of LLM},
  author={Xing, Jiarong and Qiao, Yifan and Mo, Simon and Cui, Xingqi and Sela, Gur-Eyal and Zhou, Yang and Gonzalez, Joseph and Stoica, Ion},
  journal={arXiv preprint arXiv:2508.08448},
  year={2025}
}

@article{lou2025hydraserve,
  title = {{HydraServe}: Minimizing Cold Start Latency for Serverless {LLM} Serving in Public Clouds},
  author = {Lou, Chiheng and Qi, Sheng and Jin, Chao and Nie, Dapeng and Yang, Haoran and Ding, Yu and Liu, Xuanzhe and Jin, Xin},
  journal = {arXiv preprint arXiv:2502.15524},
  year = {2025},
}

@inproceedings{narayanan2019pipedream,
  title = {{PipeDream}: Generalized Pipeline Parallelism for {DNN} Training},
  author = {Narayanan, Deepak and Harlap, Aaron and Phanishayee, Amar and Seshadri, Vivek and Devanur, Nikhil R. and Ganger, Gregory R. and Gibbons, Phillip B. and Zaharia, Matei},
  booktitle = {Proceedings of the 27th ACM Symposium on Operating Systems Principles (SOSP '19)},
  year = {2019},
}

@inproceedings{dao2022flashattention,
  title = {{FlashAttention}: Fast and Memory-Efficient Exact Attention with {IO}-Awareness},
  author = {Dao, Tri and Fu, Daniel Y. and Ermon, Stefano and Rudra, Atri and R{\'e}, Christopher},
  booktitle = {Advances in Neural Information Processing Systems (NeurIPS)},
  year = {2022},
}

@article{openai2023gpt4,
  title = {{GPT-4} Technical Report},
  author = {{OpenAI}},
  journal = {arXiv preprint arXiv:2303.08774},
  year = {2023},
}

@article{griggs2024melange,
  title = {M{\'e}lange: Cost Efficient Large Language Model Serving by Exploiting {GPU} Heterogeneity},
  author = {Griggs, Tyler and Liu, Xiaoxuan and Yu, Jiaxiang and Kim, Doyoung and Chiang, Wei-Lin and Cheung, Alvin and Stoica, Ion},
  journal = {arXiv preprint arXiv:2404.14527},
  year = {2024},
}

@article{touvron2023llama,
  title = {{LLaMA}: Open and Efficient Foundation Language Models},
  author = {Touvron, Hugo and Lavril, Thibaut and Izacard, Gautier and Martinet, Xavier and Lachaux, Marie-Anne and Lacroix, Timoth{\'e}e and Rozi{\`e}re, Baptiste and Goyal, Naman and Hambro, Eric and Azhar, Faisal and others},
  journal = {arXiv preprint arXiv:2302.13971},
  year = {2023},
}

@article{grattafiori2024llama3,
  title = {The Llama 3 Herd of Models},
  author = {Grattafiori, Aaron and Dubey, Abhimanyu and Jauhri, Abhinav and Pandey, Abhinav and Kadian, Abhishek and Al-Dahle, Ahmad and Letman, Aiesha and Mathur, Akhil and Schelten, Alan and Vaughan, Amy and others},
  journal = {arXiv preprint arXiv:2407.21783},
  year = {2024},
}

@inproceedings{yu2022orca,
  title = {{Orca}: A Distributed Serving System for Transformer-Based Generative Models},
  author = {Yu, Gyeong-In and Jeong, Joo Seong and Kim, Geon-Woo and Kim, Soojeong and Chun, Byung-Gon},
  booktitle = {Proceedings of the 16th USENIX Symposium on Operating Systems Design and Implementation (OSDI '22)},
  year = {2022},
}

@article{zheng2023alpa,
  title = {{Alpa}: Automating Inter- and Intra-Operator Parallelism for Distributed Deep Learning},
  author = {Zheng, Lianmin and Li, Zhuohan and Zhang, Hao and Zhuang, Yonghao and Chen, Zhifeng and Huang, Yanping and Wang, Yida and Xu, Yuanzhong and Zhuo, Danyang and Xing, Eric P. and Gonzalez, Joseph E. and Stoica, Ion},
  booktitle = {Proceedings of the 16th USENIX Symposium on Operating Systems Design and Implementation (OSDI '22)},
  year = {2022},
}

@article{vaswani2017attention,
  title = {Attention is All You Need},
  author = {Vaswani, Ashish and Shazeer, Noam and Parmar, Niki and Uszkoreit, Jakob and Jones, Llion and Gomez, Aidan N. and Kaiser, {\L}ukasz and Polosukhin, Illia},
  booktitle = {Advances in Neural Information Processing Systems (NeurIPS)},
  year = {2017},
}

@article{brown2020gpt3,
  title = {Language Models are Few-Shot Learners},
  author = {Brown, Tom B. and Mann, Benjamin and Ryder, Nick and Subbiah, Melanie and Kaplan, Jared and Dhariwal, Prafulla and Neelakantan, Arvind and Shyam, Pranav and Sastry, Girish and Askell, Amanda and others},
  booktitle = {Advances in Neural Information Processing Systems (NeurIPS)},
  year = {2020},
}

@inproceedings{zheng2024sglang,
  title = {{SGLang}: Efficient Execution of Structured Language Model Programs},
  author = {Zheng, Lianmin and Yin, Liangsheng and Xie, Zhiqiang and Sun, Chuyue and Huang, Jeff and Yu, Cody Hao and Cao, Shiyi and Kozyrakis, Christos and Stoica, Ion and Gonzalez, Joseph E. and Barrett, Clark and Sheng, Ying},
  booktitle = {Advances in Neural Information Processing Systems (NeurIPS)},
  year = {2024},
}

@inproceedings{he2025uellm,
  title = {{UELLM}: A Unified and Efficient Approach for Large Language Model Inference Serving},
  author = {He, Yiyuan and Xu, Minxian and Wu, Jingfeng and Zheng, Wanyi and Ye, Kejiang and Xu, Chengzhong and Gaaloul, Walid and Sheng, Michael and Yu, Qi and Yangui, Sami},
  booktitle = {International Conference on Service-Oriented Computing (ICSOC)},
  year = {2025},
}

@article{lin2025flexpipe,
  title = {{FlexPipe}: Adapting Dynamic {LLM} Serving Through Inflight Pipeline Refactoring in Fragmented Serverless Clusters},
  author = {Lin, Yanying and Peng, Shijie and Lu, Chengzhi and Xu, Chengzhong and Ye, Kejiang},
  journal = {arXiv preprint arXiv:2510.11938},
  year = {2025},
}

@article{wang2024burstgpt,
  title = {{BurstGPT}: A Real-world Workload Dataset to Optimize {LLM} Serving Systems},
  author = {Wang, Yuxin and Chen, Yuhan and Li, Zeyu and Kang, Xueze and Fang, Yuchu and Zhou, Yeju and Zheng, Yang and Tang, Zhenheng and He, Xin and Guo, Rui and Wang, Xin and Wang, Qiang and Zhou, Amelie Chi and Chu, Xiaowen},
  journal = {arXiv preprint arXiv:2401.17644},
  year = {2024},
}

@inproceedings{sun2024llumnix,
  title = {{Llumnix}: Dynamic Scheduling for Large Language Model Serving},
  author = {Sun, Biao and Huang, Ziming and Zhao, Hanyu and Xiao, Wencong and Zhang, Xinyi and Li, Yong and Lin, Wei},
  booktitle = {Proceedings of the 18th USENIX Symposium on Operating Systems Design and Implementation (OSDI '24)},
  year = {2024},
}

@inproceedings{zhong2024distserve,
  title = {{DistServe}: Disaggregating Prefill and Decoding for Goodput-optimized Large Language Model Serving},
  author = {Zhong, Yinmin and Liu, Shengyu and Chen, Junda and Hu, Jianbo and Zhu, Yibo and Liu, Xuanzhe and Jin, Xin and Zhang, Hao},
  booktitle = {Proceedings of the 18th USENIX Symposium on Operating Systems Design and Implementation (OSDI '24)},
  year = {2024},
}

@inproceedings{fu2024serverlessllm,
  title = {{ServerlessLLM}: Low-Latency Serverless Inference for Large Language Models},
  author = {Fu, Yao and Xue, Leyang and Huang, Yeqi and Brabete, Andrei-Octavian and Ustiugov, Dmitrii and Patel, Yuvraj and Mai, Luo},
  booktitle = {Proceedings of the 18th USENIX Symposium on Operating Systems Design and Implementation (OSDI '24)},
  year = {2024},
}

@inproceedings{miao2024spotserve,
  title = {{SpotServe}: Serving Generative Large Language Models on Preemptible Instances},
  author = {Miao, Xupeng and Shi, Chunan and Duan, Jiangfei and Xi, Xiaoli and Lin, Dahua and Cui, Bin and Jia, Zhihao},
  booktitle = {Proceedings of the 29th ACM International Conference on Architectural Support for Programming Languages and Operating Systems, Volume 2 (ASPLOS '24)},
  pages = {1112--1127},
  year = {2024},
  doi = {10.1145/3620665.3640411},
}

@inproceedings{li2023lyra,
  title = {{Lyra}: Elastic Scheduling for Deep Learning Clusters},
  author = {Li, Jiamin and Xu, Hong and Zhu, Yibo and Liu, Zherui and Guo, Chuanxiong and Wang, Cong},
  booktitle = {Proceedings of the Eighteenth European Conference on Computer Systems (EuroSys '23)},
  pages = {835--850},
  year = {2023},
  doi = {10.1145/3552326.3587445},
}

@inproceedings{suSeesawHighthroughputLLM2025,
  title = {Seesaw: {{High-throughput LLM Inference}} via {{Model Re-sharding}}},
  shorttitle = {Seesaw},
  author = {Su, Qidong and Zhao, Wei and Li, Xin and Andoorveedu, Muralidhar and Jiang, Chenhao and Zhu, Zhanda and Song, Kevin and Giannoula, Christina and Pekhimenko, Gennady},
  booktitle = {Proceedings of Machine Learning and Systems (MLSys)},
  year = {2025}
}

@misc{yuanLLMInferenceUnveiled2024,
  title = {{{LLM Inference Unveiled}}: {{Survey}} and {{Roofline Model Insights}}},
  author = {Yuan, Zhihang and Shang, Yuzhang and Zhou, Yang and Dong, Zhen and Zhou, Zhe and Xue, Chenhao and Wu, Bingzhe and Li, Zhikai and Gu, Qingyi and Lee, Yong Jae and Yan, Yan and Chen, Beidi and Sun, Guangyu and Keutzer, Kurt},
  year = {2024},
  eprinttype = {arXiv},
  eprintclass = {cs},
}

@inproceedings{qinMooncake2025,
  title = {Mooncake: Trading More Storage for Less Computation --- A {KVCache}-Centric Architecture for Serving {LLM} Chatbot},
  author = {Qin, Ruoyu and Li, Zheming and He, Weiran and Cui, Junda and Ren, Fangcheng and Zhang, Mingxing and Wu, Yongwei and Zheng, Weimin and Xu, Xinran},
  booktitle = {23rd USENIX Conference on File and Storage Technologies (FAST 25)},
  year = {2025}
}

@inproceedings{moHetis2025,
  title = {Hetis: Serving LLMs in Heterogeneous GPU Clusters with Fine-grained and Dynamic Parallelism},
  author = {Mo, Zizhao and Liao, Jianxiong and Xu, Huanle and Zhou, Zhi and Xu, Chengzhong},
  booktitle = {Proceedings of the International Conference for High Performance Computing, Networking, Storage, and Analysis (SC)},
  year = {2025},
}

@inproceedings{nvidiaNCCL,
  title = {Demystifying {NCCL}: An In-Depth Analysis of {GPU} Communication Protocols and Algorithms},
  author = {Hu, Zixuan and Shen, Siyuan and Bonato, Tommaso and Jeaugey, Sylvain and Hoefler, Torsten},
  booktitle = {Proceedings of the 39th IEEE International Parallel and Distributed Processing Symposium (IPDPS)},
  year = {2025},
}

@inproceedings{clark2005live,
  title = {Live Migration of Virtual Machines},
  author = {Clark, Christopher and Fraser, Keir and Hand, Steven and Hansen, Jacob Gorm and Jul, Eric and Limpach, Christian and Pratt, Ian and Warfield, Andrew},
  booktitle = {Proceedings of the 2nd Conference on Symposium on Networked Systems Design \& Implementation (NSDI)},
  year = {2005},
}

@inproceedings{huang2019gpipe,
  title = {{GPipe}: Efficient Training of Giant Neural Networks Using Pipeline Parallelism},
  author = {Huang, Yanping and Cheng, Youlong and Bapna, Ankur and Firat, Orhan and Chen, Mia Xu and Chen, Dehao and Lee, HyoukJoong and Ngiam, Jiquan and Le, Quoc V. and Wu, Yonghui and Chen, Zhifeng},
  booktitle = {Advances in Neural Information Processing Systems (NeurIPS)},
  year = {2019}
}

@inproceedings{dao2023flashattention2,
  title = {{FlashAttention-2}: Faster Attention with Better Parallelism and Work Partitioning},
  author = {Dao, Tri},
  booktitle = {International Conference on Learning Representations (ICLR)},
  year = {2024}
}

@inproceedings{agrawal2024sarathiserve,
  title = {Taming Throughput-Latency Tradeoff in {LLM} Inference with {Sarathi-Serve}},
  author = {Agrawal, Amey and Kedia, Nitin and Panwar, Ashish and Mohan, Jayashree and Kwatra, Nipun and Gulavani, Bhargav S. and Ramjee, Ramachandran and Tumanov, Alexey},
  booktitle = {18th USENIX Symposium on Operating Systems Design and Implementation (OSDI)},
  year = {2024}
}

@inproceedings{prabhu2024vattention,
  title = {{vAttention}: Dynamic Memory Management for Serving {LLMs} without {PagedAttention}},
  author = {Prabhu, Ramya and Nayak, Ajay and Mohan, Jayashree and Ramjee, Ramachandran and Panwar, Ashish},
  booktitle = {Proceedings of the ACM SIGOPS 30th Symposium on Operating Systems Principles (SOSP)},
  year = {2024},
}

@inproceedings{xudynapipe,
  title={DynaPipe: Dynamic Layer Redistribution for Efficient Serving of LLMs with Pipeline Parallelism},
  author={Xu, Hongxin and Guo, Tianyu and Zhang, Xianwei},
  booktitle={The Thirty-ninth Annual Conference on Neural Information Processing Systems}
}

@inproceedings{wagenlander2024tenplex,
  title = {Tenplex: Dynamic Parallelism for Deep Learning Using Parallelizable Tensor Collections},
  author = {Wagenl{\"a}nder, Marcel and Li, Guo and Zhao, Bo and Mai, Luo and Pietzuch, Peter},
  booktitle = {Proceedings of the ACM SIGOPS 30th Symposium on Operating Systems Principles (SOSP)},
  year = {2024},
}

\end{document}